\documentclass[%
 reprint,
 amsmath,amssymb,
 aps,
 prl,
 longbibliography
]{revtex4-2}

\usepackage[pdftex]{graphicx} % for arXiv
\usepackage{dcolumn}% Align table columns on decimal point
\usepackage{bm}% bold math
\usepackage[T1]{fontenc}
\usepackage[pdftex]{color, hyperref}% for arXiv add hypertext capabilities
\hypersetup{
colorlinks = true,
linkcolor = blue,
citecolor = blue
}
\usepackage{here} %図Here位置強制
\usepackage{color}

\newcommand{\parhyphen}[1][1em]{% Symmetric \paragraph hyphen
  {\normalfont\normalsize\bfseries\hspace*{-1em}}\hspace*{#1}---\hspace*{#1}\ignorespaces}

\begin{document}
\footnotetext[1]{See Supplemental Material at [URL will be inserted by publisher] for details on the derivation and numerical simulation.}

% \preprint{APS/123-QED}

\title{Scale-to-Scale Information Flow Amplifies Turbulent Fluctuations}% Force line breaks with \\

\author{Tomohiro Tanogami$^1$}
\author{Ryo Araki$^{2}$}
\affiliation{$^1$Department of Earth and Space Science, Osaka University, Osaka 560-0043, Japan\\
$^2$Department of Mechanical and Aerospace Engineering, Faculty of Science
and Technology, Tokyo University of Science, Yamazaki 2641, Noda-shi
278-8510, Japan}%Lines break automatically or can be forced with \\

% \collaboration{MUSO Collaboration}%\noaffiliation

% \affiliation{
% \affiliation{
% \affiliation{%
%  Authors' institution and/or address\\
%  This line break forced with \textbackslash\textbackslash
% }%

% \collaboration{CLEO Collaboration}%\noaffiliation

\date{\today}% It is always \today, today,
\begin{abstract}
% \begin{description}
% Secondary publications and information retrieval purposes.
% \item[PACS numbers]
% May be entered using the \verb+\pacs{#1}+ command.
% You may use the \texttt{description} environment to structure your abstract;
% use the optional argument of the \verb+\item+ command to give the category of each item.
% \end{description}
%%%%%% Introduction
In three-dimensional turbulence, information of turbulent fluctuations at large scales is propagated to small scales.
%%%%%% Details of motivation/previous studies
%%%%%% Research question
Here, we investigate the relation between the information flow and turbulent fluctuations described by a shell model.
%%%%%% Result
We first establish a connection between the information flow and phase-space contraction rate.
From this relation, we then prove an inequality between the information flow and turbulent fluctuations, which suggests that the information flow from large to small scales amplifies turbulent fluctuations at small scales.
%%%%%% Details of the result
%%%%%% Significance/impact
%%%%%% Implications
This inequality can also be interpreted as a quantification of Landau's objection to the universality of turbulent fluctuations.
We also discuss differences between the information flow and the Kolmogorov--Sinai entropy.
\end{abstract}

\pacs{Valid PACS appear here}% PACS, the Physics and Astronomy

\maketitle
\paragraph*{Introduction.}\parhyphen[0pt]
%%%%%% Background: Universal statistics of turbulent fluctuations
Turbulent fluctuations intrinsically affect the accuracy of prediction and control of various complex flow phenomena observed, for example, in the Earth system~\cite{lorenz1969predictability,palmer2014real,palmer2024real}.
To elucidate the fundamental bounds on the prediction and control of these phenomena, we must clarify the nature of turbulent fluctuations.
One of the most prominent properties of turbulent fluctuations is a universal scaling law.
For example, in three-dimensional turbulence, the $p$th-order moment of the longitudinal velocity increment across a distance $\ell$ defined by $\delta u_{\parallel}({\bm \ell};{\bm x}):=({\bm u}({\bm x}+{\bm \ell})-{\bm u}({\bm x}))\cdot{\bm \ell}/\ell$ exhibits power-law scaling, $\langle(\delta u_{\parallel}(\ell))^p\rangle\propto\ell^{\zeta_p}$, with a scaling exponent $\zeta_p$~\cite{Frisch,bohr1998dynamical,davidson2015turbulence,Eyink_lecture}.
This scaling law holds for $\ell$ in the inertial range, where both the forcing and viscous effects are negligible and energy cascades from large to small scales.
While the magnitude of the turbulent fluctuation itself may not be universal, as Landau pointed out in Ref.~\cite{landau1959fluid} (see also Refs.~\cite{kraichnan1974kolmogorov,Frisch}), the scaling exponent $\zeta_p$ is believed to be universal in the sense that it is independent of the details of large-scale statistics.
Despite numerous theoretical attempts, the scaling exponents have never been rigorously obtained except for $p=3$, and their analytical calculation is referred to as the \textit{Holy Grail} of turbulence~\cite{de2024extreme}.
Because the scaling law emerges from the interplay between turbulent fluctuations over a wide range of scales, it is desirable to elucidate the underlying constraints governing the interference of turbulent fluctuations.

%%%%%% Background: Universal bounds elucidated from the viewpoint of information
Various constraints inherent in the dynamics of strongly interacting classical and quantum many-body systems can be elucidated from the perspective of information theory~\cite{cover1999elements}.
In particular, \textit{information thermodynamics} provides deep insights into the connection between information and thermodynamics and has made it possible to reveal a variety of universal bounds on the dynamics of many-body systems~\cite{peliti2021stochastic,shiraishi2023introduction,parrondo2015thermodynamics,horowitz2014thermodynamics,horowitz2015multipartite,ehrich2023energy,tanogami2023universal,goold2016role,gong2022bounds,tanogami2023universal}.
While there have been several attempts to apply information theory to turbulence~\cite{betchov1964measure,ikeda1989information,cerbus2013information,materassi2014information,cerbus2016information,goldburg2016turbulence,granero2016scaling,granero2018kullback,lozano2020causality,shavit2020singular,vladimirova2021fibonacci,PhysRevResearch.4.023195,arranz2024informative,araki2024forgetfulness,yatomi2024quantum}, these previous studies have mainly focused on quantifying causality and statistical properties by numerically estimating information-theoretic quantities.
In contrast, we aim to elucidate universal constraints on turbulence dynamics from an information-thermodynamic viewpoint.

%%%%%% Background: Information flow in turbulence
%%%%%% Research question & Approach
In our previous study, we applied information thermodynamics to turbulence and proved that the information of turbulent fluctuations at large scales is propagated to small scales in the inertial range~\cite{tanogami2024information}.
The fact that there is an information transfer from large to small scales suggests that turbulent fluctuations at small scales are constrained by the information flow.
The purpose of this Letter is to investigate the universal relation between the information flow and turbulent fluctuations.

As a first step to this end, we focus on a shell model.
We first establish a connection between the information flow and phase-space contraction rate, a dynamical quantity that has been studied in detail in relation to the entropy production rate and fluctuation relations~\cite{evans2002fluctuation,rondoni2007fluctuations,baiesi2015inflow,gallavotti2020nonequilibrium}.
From this relation, we prove an inequality between the information flow and turbulent fluctuations, which suggests that the information flow amplifies turbulent fluctuations.
This inequality can also be interpreted as a quantification of Landau's objection to the universality of turbulent fluctuations~\cite{Frisch,kraichnan1974kolmogorov,landau1959fluid}, suggesting that the magnitude of turbulent fluctuations is not universal but depends on large-scale statistics through the information flow.
These findings also highlight differences between the information flow and the Kolmogorov--Sinai (KS) entropy, which quantifies the rate of information loss in chaotic dynamics~\cite{eckmann1985ergodic,boffetta2002predictability,bohr1998dynamical,ruelle1982large,ruelle1984characteristic,berera2019information}.

\paragraph*{Setup.}\parhyphen[0pt]
We consider the Sabra shell model with thermal noise~\cite{l1998improved,bandak2021thermal,bandak2022dissipation}, which is a simplified model of the fluctuating Navier--Stokes equation~\cite{landau1959fluid,de2006hydrodynamic}.
Although we include thermal noise to clarify the connection with our previous studies~\cite{tanogami2024information,tanogami2024scale}, the same results can also be obtained for the deterministic case.

Let $u_n(t)\in\mathbb{C}$ be the ``velocity'' at time $t$ with wavenumber $k_n=k_02^n$ ($n=0,1,\cdots,N$).
The time evolution of the shell variables $u:=\{u_n\}$ is governed by the following Langevin equation~\cite{bandak2021thermal,bandak2022dissipation}:
\begin{align}
\partial_t{u}_n=B_n(u,u^*)-\nu k^2_nu_n+\sqrt{\dfrac{2\nu k^2_nk_{\mathrm{B}}T}{\rho}}\xi_n+f_n.
\label{sabra shell model}
\end{align}
Here, $B_n(u,u^*)$ denotes the scale-local nonlinear interactions defined by
\begin{align}
B_n(u,u^*)&:=i\biggl(k_{n+1}u_{n+2}u^*_{n+1}-\dfrac{1}{2}k_nu_{n+1}u^*_{n-1}\biggr.\notag\\
&\qquad\quad\biggl.+\dfrac{1}{2}k_{n-1}u_{n-1}u_{n-2}\biggr)
\end{align}
with $u_{-1}=u_{-2}=u_{N+1}=u_{N+2}=0$, $\nu$ represents ``kinematic viscosity'', and $f_n\in\mathbb{C}$ denotes the external body force that acts only at large scales, i.e., $f_n=0$ for $n>n_f$.
The third term on the right-hand side of Eq.~(\ref{sabra shell model}) denotes the thermal noise, where $\xi_n\in\mathbb{C}$ is the zero-mean white Gaussian noise that satisfies $\langle\xi_n(t)\xi^*_{n'}(t')\rangle=2\delta_{nn'}\delta(t-t')$, $T$ denotes the absolute temperature, $k_{\mathrm{B}}$ the Boltzmann constant, and $\rho$ the mass ``density''.
The strength of the noise satisfies the fluctuation-dissipation relation of the second kind~\cite{maes2021local,peliti2021stochastic,shiraishi2023introduction}.

Let $p_t(u,u^*)$ be the probability distribution of state $(u,u^*)$ at time $t$.
The time evolution of $p_t(u,u^*)$ is governed by the Fokker--Planck equation~\cite{risken1996fokker,gardiner1985handbook}:
\begin{align}
\partial_tp_t(u,u^*)=\sum^N_{n=0}\left[-\dfrac{\partial}{\partial u_n}J_n(u,u^*)-\dfrac{\partial}{\partial u^*_n}J^*_n(u,u^*)\right],
% &=\sum^N_{n=0}\left[-\dfrac{\partial}{\partial u_n}\left(A_n(u,u^*)p_t(u,u^*)\right)-\dfrac{\partial}{\partial u^*_n}\left(A^*_n(u,u^*)p_t(u,u^*)\right)+\dfrac{4\nu k^2_nk_{\mathrm{B}}T}{\rho}\dfrac{\partial^2}{\partial u_n\partial u^*_n}p_t(u,u^*)\right]\notag\\
\label{FP-sabra}
\end{align}
where $J_n(u,u^*)$ denotes the probability current associated with the shell variable $u_n$:
\begin{align}
J_n(u,u^*)&:=\left(B_n(u,u^*)-\nu k^2_nu_n+f_n\right)p_t(u,u^*)\notag\\
&\qquad-\dfrac{2\nu k^2_nk_{\mathrm{B}}T}{\rho}\dfrac{\partial}{\partial u^*_n}p_t(u,u^*).
\label{probability current}
\end{align}

This model is known to exhibit rich temporal and multiscale statistics that are similar to those observed in real turbulent flow~\cite{bohr1998dynamical,l1998improved,biferale2003shell,bandak2022dissipation}.
For example, the scale-local nonlinear interactions cause the energy cascade in the inertial range $k_f\ll k_n\ll k_\nu$, where $k_f:=k_{n_f}$ denotes the energy injection scale and $k_\nu:=\nu^{-3/4}\varepsilon^{1/4}$ denotes the energy dissipation scale.
Here, $\varepsilon$ denotes the energy dissipation rate defined by $\varepsilon:=\sum^N_{n=0}\nu k^2_n\langle|u_n|^2\rangle$.
Along with the energy cascade, the $p$th-order velocity structure functions $\langle|u_n|^p\rangle$ exhibit universal scaling laws $\propto k^{-\zeta_p}_n$ with nonlinear scaling exponents $\zeta_p$ similar to those of real fluid turbulence~\cite{de2024extreme}.

\begin{figure}[t]
\includegraphics[width=8.6cm]{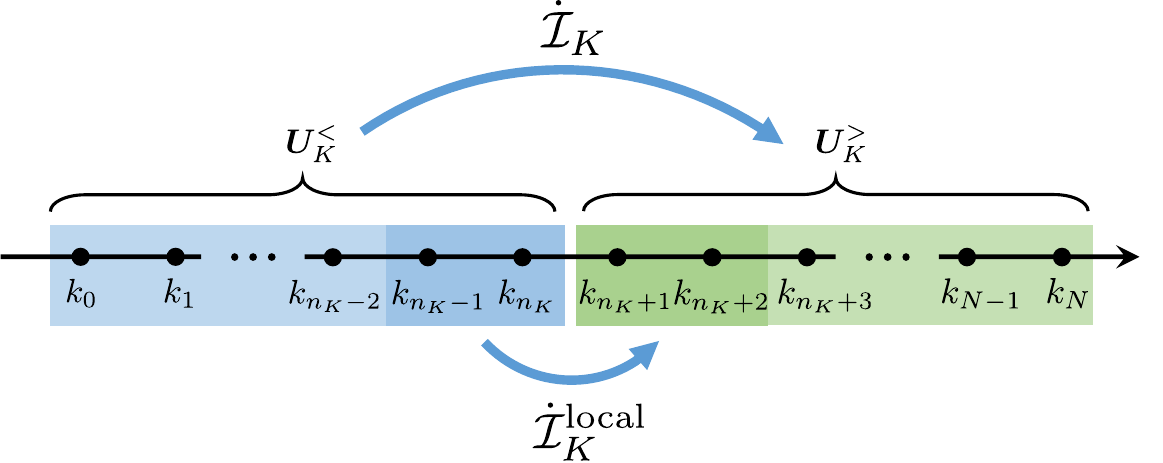}
\caption{Schematic of the scale-to-scale information flow $\dot{\mathcal{I}}_K$ (the upper arrow). The lower arrow indicates the scale-local information flow $\dot{\mathcal{I}}^{\mathrm{local}}_K$, which quantifies the information transfer from the dark blue region to the dark green region.}
\label{fig:Fourier_modes_division}
\end{figure}

Now, we introduce the \textit{scale-to-scale information flow}~\cite{tanogami2024information,tanogami2024scale}.
We first divide the total shell variables $\{u,u^*\}$ into two parts at an arbitrary intermediate scale $K:=k_{n_K}$ with $n_K\in\{0,\cdots,N\}$ as $\{u,u^*\}={\bm U}^<_K\cup{\bm U}^>_K$, where ${\bm U}^<_K:=\{u_n,u^*_n\mid0\le n\le n_K\}$ and ${\bm U}^>_K:=\{u_n,u^*_n\mid n_K+1\le n\le N\}$ denote the large-scale and small-scale modes, respectively (see Fig.~\ref{fig:Fourier_modes_division}).

The strength of the correlation between the large-scale modes ${\bm U}^<_K$ and small-scale modes ${\bm U}^>_K$ at time $t$ is quantified by \textit{mutual information}~\cite{cover1999elements}:
\begin{align}
% I[{\bm U}^<_K\colon\!{\bm U}^>_K]:=\int d{\bm U}^<_Kd{\bm U}^>_Kp_t({\bm U}^<_K,{\bm U}^>_K)\ln\dfrac{p_t({\bm U}^<_K,{\bm U}^>_K)}{p_t({\bm U}^<_K)p_t({\bm U}^>_K)},
I[{\bm U}^<_K(t)\colon\!{\bm U}^>_K(t)]:=\left\langle\ln\dfrac{p_t({\bm U}^<_K,{\bm U}^>_K)}{p_t({\bm U}^<_K)p_t({\bm U}^>_K)}\right\rangle,
\label{def: MI}
\end{align}
where $\langle\cdot\rangle$ denotes the average with respect to the joint probability distribution $p_t({\bm U}^<_K,{\bm U}^>_K)=p_t(u,u^*)$, and $p_t({\bm U}^<_K)$ and $p_t({\bm U}^>_K)$ are the marginal distributions for the large-scale and small-scale modes, respectively.
The mutual information is nonnegative and is equal to zero if and only if ${\bm U}^<_K$ and ${\bm U}^>_K$ are statistically independent.

The flow of information can be quantified by the time derivative of the mutual information, which is called \textit{information flow}~\cite{allahverdyan2009thermodynamic,hartich2014stochastic,barato2014efficiency,hartich2016sensory,matsumoto2018role}.
Although there are many other quantities that describe the flow of information, such as \textit{transfer entropy}~\cite{schreiber2000measuring,materassi2014information} and \textit{information flux}~\cite{PhysRevResearch.4.023195,araki2024forgetfulness}, this information flow is appropriate for our argument because it appears in the second law of information thermodynamics~\cite{horowitz2014thermodynamics,ehrich2023energy} and plays a crucial role in understanding the universal constraint on dynamics.
We define the \textit{scale-to-scale information flow} $\dot{I}^>_K$ at scale $K$ as the information flow from ${\bm U}^<_K$ to ${\bm U}^>_K$:
\begin{align}
\dot{I}^>_K:=\lim_{dt\rightarrow0^+}\dfrac{I[{\bm U}^<_K(t)\colon\!{\bm U}^>_K(t+dt)]-I[{\bm U}^<_K(t)\colon\!{\bm U}^>_K(t)]}{dt}.
% &=\sum^{N}_{n=n_K+1}\int dudu^*\left[J_n(u,u^*)\dfrac{\partial}{\partial u_n}\ln\dfrac{p_t(u,u^*)}{p_t({\bm U}^<_K)p_t({\bm U}^>_K)}\right.\notag\\
% &\qquad\qquad\left.+J^*_n(u,u^*)\dfrac{\partial}{\partial u^*_n}\ln\dfrac{p_t(u,u^*)}{p_t({\bm U}^<_K)p_t({\bm U}^>_K)}\right].
\label{IF_def-2}
\end{align}
The scale-to-scale information flow $\dot{I}^<_K$ from ${\bm U}^>_K$ to ${\bm U}^<_K$ can also be defined in a similar manner.
It then immediately follows that $d_tI[{\bm U}^<_K\colon\!{\bm U}^>_K]=\dot{I}^<_K+\dot{I}^>_K$~\cite{tanogami2024information,tanogami2024scale}.
We introduce the notation $\dot{\mathcal{I}}_K$ to denote the time-independent steady-state information flow $\dot{I}^>_K$, which satisfies $\dot{\mathcal{I}}_K=\dot{I}^>_K=-\dot{I}^<_K$.
If $\dot{\mathcal{I}}_K>0$ ($\dot{\mathcal{I}}_K<0$), then it means that information of turbulent fluctuations is transferred from large (small) to small (large) scales (see Fig.~\ref{fig:Fourier_modes_division}). 
In our previous paper~\cite{tanogami2024information}, by applying information thermodynamics, we proved that $\dot{\mathcal{I}}_K\ge0$ for $k_f\ll K\ll k_\nu$.

\paragraph*{Effective large-scale dynamics.}\parhyphen[0pt]
Here, we formulate the effective large-scale dynamics of the shell model (\ref{sabra shell model}) following Ref.~\cite{biferale2017optimal}.
We set the cutoff wavenumber $K=k_{n_K}$ within the inertial range.
By integrating out the Fokker--Planck equation (\ref{FP-sabra}) with the small-scale modes ${\bm U}^>_K$, we obtain the time evolution equation of the marginal distribution for the large-scale modes:
\begin{align}
\partial_tp_t({\bm U}^<_K)=\sum^{n_K}_{n=0}\left[-\dfrac{\partial}{\partial u_n}\overline{J}_n({\bm U}^<_K)-\dfrac{\partial}{\partial u^*_n}\overline{J}^*_n({\bm U}^<_K)\right].
\label{Fokker-Planck effective}
\end{align} 
Here, $\overline{J}_n({\bm U}^<_K):=\left(\overline{B}_n({\bm U}^<_K)+f_n\right)p_t({\bm U}^<_K)$ denotes the effective probability current, where we ignored the viscous and thermal noise terms, and introduced the effective nonlinear term $\overline{B}_n({\bm U}^<_K)$ as the conditional average of $B_n(u,u^*)$ with respect to the conditional probability density $p_t({\bm U}^>_K|{\bm U}^<_K):=p_t(u,u^*)/p_t({\bm U}^<_K)$:
\begin{align}
\overline{B}_n({\bm U}^<_K):=\int d{\bm U}^>_KB_n(u,u^*)p_t({\bm U}^>_K|{\bm U}^<_K).
\label{effective nonlinear term}
\end{align} 
The reduced Fokker--Planck equation (\ref{Fokker-Planck effective}) describes the effective large-scale dynamics:
\begin{align}
\partial_tu_n=\overline{B}_n({\bm U}^<_K)+f_n\quad\text{for}\quad 0\le n\le n_K.
\end{align}

Note that $\overline{B}_n({\bm U}^<_K)$ depends on the time evolution of $p_t({\bm U}^>_K|{\bm U}^<_K)$, and thus, Eq.~(\ref{Fokker-Planck effective}) is not closed.
However, by assuming Kolmogorov's hypothesis for the Kolmogorov multiplier~\cite{kolmogorov1962refinement}, we can show that $\overline{B}_n({\bm U}^<_K)$ is a universal function independent of $t$.
In the shell model, the multipliers $z_n\in\mathbb{C}$ are defined by $z_n:=|u_n/u_{n-1}|e^{i\Delta_n}$, where $\Delta_n:=\theta_n-\theta_{n-1}-\theta_{n-2}$ with $\theta_n:=\arg u_n$~\cite{benzi1993intermittency,eyink2003gibbsian,biferale2017optimal}.
There is a one-to-one correspondence between the multipliers $\{z_n\}$ and $\{u_n\}$.
Note that $z_0$ is not defined because $u_{-1}=0$.
Then, Kolmogorov's hypothesis states that the single-time statistics of multipliers is universal and independent of the shell number $n$ in the inertial range and that the multipliers for widely separated shells are statistically independent.
This hypothesis is confirmed numerically and experimentally~\cite{eyink2003gibbsian,chen2003kolmogorov,benzi1998multiscale,pedrizzetti1996self,chhabra1992scale}.
Under this assumption, $p_t({\bm U}^>_K|{\bm U}^<_K)$ can be expressed as
\begin{align}
p_t({\bm U}^>_K|{\bm U}^<_K)d{\bm U}^>_K=p_{\mathrm{uni}}({\bm Z}^>_K|{\bm Z}^{<,\mathrm{local}}_K)d{\bm Z}^>_K,
\label{Kolmogorov hypothesis}
\end{align}
where ${\bm Z}^>_K:=\{z_n,z^*_n\mid n_K+1\le n\le N\}$ denotes the small-scale multipliers, and $p_{\mathrm{uni}}({\bm Z}^>_K|{\bm Z}^{<,\mathrm{local}}_K)$ denotes the conditional probability density for ${\bm Z}^>_K$, which is universal and time-independent in the developed turbulent regime~\cite{biferale2017optimal}.
Here, we used the notation ${\bm Z}^{<,\mathrm{local}}_K:=\{z_{n_K},z^*_{n_K},z_{n_K-1},z^*_{n_K-1},\ldots\}$ to indicate that $p_{\mathrm{uni}}({\bm Z}^>_K|{\bm Z}^{<,\mathrm{local}}_K)$ has a weak dependence on $z_n$ for $n\ll n_K$.
The relation (\ref{Kolmogorov hypothesis}) implies that $\overline{B}_n({\bm U}^<_K)$ is universal, and that Eq.~(\ref{Fokker-Planck effective}) is closed with respect to $p_t({\bm U}^<_K)$.

While the divergence of $B_n(u,u^*)$ for each shell is zero (the Liouville theorem), i.e.,
\begin{align}
\dfrac{\partial}{\partial u_n}B_n(u,u^*)+\dfrac{\partial}{\partial u^*_n}B^*_n(u,u^*)=0,
\label{Liouville theorem}
\end{align}
the divergence of $\overline{B}_n({\bm U}^<_K)$ is generally nonzero.
In fact, $\overline{B}_n({\bm U}^<_K)$ can be interpreted as including the turbulent eddy viscosity~\cite{biferale2017optimal}.
The eddy viscosity induces a contraction of the phase-space volumes, which is quantified by~\cite{evans2002fluctuation,rondoni2007fluctuations,baiesi2015inflow,gallavotti2020nonequilibrium}
\begin{align}
\dot{\sigma}_K:=-\left\langle\sum^{n_K}_{n=0}\left(\dfrac{\partial}{\partial u_n}\overline{B}_n({\bm U}^<_K)+\dfrac{\partial}{\partial u^*_n}\overline{B}^*_n({\bm U}^<_K)\right)\right\rangle.
% :=-\sum^{n_K}_{n=0}\int d{\bm U}^<_K\left(\dfrac{\partial}{\partial u_n}\overline{B}_n({\bm U}^<_K)+\dfrac{\partial}{\partial u^*_n}\overline{B}^*_n({\bm U}^<_K)\right)p_t({\bm U}^<_K).
\label{average phase-space contraction rate}
\end{align}
% where $\langle\cdot\rangle$ denotes the average with respect to $p_t({\bm U}^<_K)$.
Note that $\dot{\sigma}_K$ denotes the average phase-space contraction rate for the large-scale modes and is different from that of the original system (\ref{sabra shell model}), which is given by $\sum_n\nu k^2_n$.

\paragraph*{Main results.}\parhyphen[0pt]
Here, we describe the main results.
The proof is provided at the end of this Letter (see also Ref.~\cite{Note1}).
The first main result is the equality between $\dot{\mathcal{I}}_K$ and $\dot{\sigma}_K$:
\begin{align}
\dot{\mathcal{I}}_K=\dot{\sigma}_K\quad\text{for}\quad k_f\ll K\ll k_\nu.
\label{first main result}
\end{align}
Equality (\ref{first main result}) states that the scale-to-scale information flow is equivalent to the average contraction rate of the phase-space volumes for large-scale modes.

An important implication of the equality (\ref{first main result}) is that the information flow is different from the KS entropy.
The KS entropy quantifies the rate of information loss in chaotic systems and is upper bounded by the sum of all the positive Lyapunov exponents~\cite{ruelle1978inequality,eckmann1985ergodic,boffetta2002predictability,bohr1998dynamical}.
In contrast, the average phase-space contraction rate is given by the negative sum of all Lyapunov exponents.
Therefore, the information flow is equal to the negative sum of all Lyapunov exponents for the effective large-scale dynamics and differs from the KS entropy of both the original and reduced dynamics.

From the first main result and Kolmogorov's hypothesis, we can further prove that $\dot{\mathcal{I}}_K$ is upper bounded by the $p$th-order velocity structure function $\langle|u_{n_K}|^p\rangle$:
\begin{align}
\dot{\mathcal{I}}_K\le C_pK\left\langle|u_{n_K}|^p\right\rangle^{1/p}
\label{IF upper bound}
\end{align}
for $p\in[1,\infty]$ and $k_f\ll K\ll k_\nu$, where $C_p$ is a universal constant defined by Eq.~(\ref{def C_p}).
The inequality (\ref{IF upper bound}) is the second main result.

Several remarks regarding the second main result are in order.
First, Eq.~(\ref{IF upper bound}) suggests that the information flow from large to small scales amplifies turbulent fluctuations at small scales.
In other words, turbulent fluctuations are influenced by the information flow.
More importantly, because $C_p$ is a universal constant, Eq.~(\ref{IF upper bound}) suggests that the magnitude of the turbulent fluctuation is not universal but depends on large-scale statistics through the information flow.
Thus, this inequality can be interpreted as a quantification of Landau's objection to the universality of turbulent fluctuations, which states that small-scale turbulent fluctuations cannot be universal because of the variation of the energy dissipation rate at large scales~\cite{landau1959fluid,kraichnan1974kolmogorov,Frisch}.
Note that this relation between the information flow and fluctuation is in contrast to that found in information processing systems with a negative feedback loop, such as biochemical signal transduction, where the information flow suppresses intrinsic fluctuations~\cite{ito2015maxwell}.

Second, although it is difficult to explicitly calculate the universal constant $C_p$, if we additionally assume statistical independence of the Kolmogorov multipliers, then $C_p$ can be evaluated as $C_p=2+2^{-2/3}=2.63\ldots=:C$, which is independent of $p$.
Furthermore, under this assumption, we can show that the equality of Eq.~(\ref{IF upper bound}) is achieved for $p=1$, i.e., $\dot{\mathcal{I}}_K=C\tau^{-1}_K$, where $\tau_K:=(K\langle|u_{n_K}|\rangle)^{-1}$ denotes the eddy turnover time at scale $K$.
For the derivation, see Supplemental Material~\cite{Note1}.

Third, Eq.~(\ref{IF upper bound}) suggests that the information flow has power-law scaling $\dot{\mathcal{I}}_K\propto k^{\alpha}$ in the inertial range with a scaling exponent $\alpha$.
Indeed, by noting that $\left\langle|u_n|^p\right\rangle^{1/p}$ follows a power-law scaling $\propto k^{-\sigma_p}_n$ with a scaling exponent $\sigma_p:=\zeta_p/p$ in the inertial range, we obtain $\alpha\le1-\sigma_p$.
Furthermore, because the exponent $\sigma_p$ is non-increasing in $p$~\cite{Eyink_lecture}, we obtain $\alpha\le1-\zeta_1\simeq2/3$.

Finally, Eq.~(\ref{IF upper bound}) has a form similar to that of the bound on the KS entropy $h_{\mathrm{KS}}$ for the Navier--Stokes equation derived by Ruelle~\cite{ruelle1982large,ruelle1984characteristic}, which reads
\begin{align}
h_{\mathrm{KS}}\le C\nu^{-11/4}\left\langle\int d{\bm r}\varepsilon^{5/4}({\bm r})\right\rangle,
\label{Ruelle bound}
\end{align}
where $C$ is a universal constant, $\langle\cdot\rangle$ denotes the average with respect to an invariant probability measure, $\varepsilon({\bm r})$ denotes the energy dissipation rate.
If we ignore the intermittency effects, the upper bound of Eq.~(\ref{Ruelle bound}) can be evaluated as $C\mathrm{Re}^{11/4}\tau^{-1}_L$, where $\mathrm{Re}$ denotes the Reynolds number and $\tau_L:=(k_0u_{\mathrm{rms}})^{-1}$ denotes the large-eddy turnover time~\cite{berera2019information}.
Note that, while the KS entropy may diverge in the infinite $\mathrm{Re}$ limit~\cite{berera2019information}, our inequality (\ref{IF upper bound}) implies that the information flow remains finite.

\paragraph*{Numerical simulation.}\parhyphen[0pt]
Here, we numerically illustrate the result (\ref{IF upper bound}).
We use the parameter values appropriate for the atmospheric boundary layer~\cite{bandak2022dissipation,tanogami2024information}.
Because it is numerically demanding to estimate $\dot{\mathcal{I}}_K$ with high precision, we instead estimate the scale-local information flow $\dot{\mathcal{I}}^{\mathrm{local}}_K$ introduced in Ref.~\cite{tanogami2024scale} (see Fig.~\ref{fig:Fourier_modes_division}).
The scale locality of the information flow proved in Ref.~\cite{tanogami2024scale} ensures that the estimated $\dot{\mathcal{I}}^{\mathrm{local}}_K$ is approximately equal to $\dot{\mathcal{I}}_K$ in the inertial range.
We estimate $\dot{\mathcal{I}}^{\mathrm{local}}_K$ using a finite difference approximation.
Specifically, we estimate the difference of mutual information $\Delta I$ with a time increment $\Delta t$ by using the Kraskov-St\"ogbauer-Grassberger (KSG) estimator~\cite{kraskov2004estimating,khan2007relative,holmes2019estimation}.
Note that the estimation error is amplified for a smaller $\Delta t$.
Because we are interested in the behavior of the information flow in the inertial range, we choose $\Delta t=0.1\tau_\nu$, where $\tau_\nu:=(k_\nu \varepsilon^{1/4}\nu^{1/4})^{-1}$ denotes the characteristic time scale at $k_\nu$.
Further details of the numerical simulations are provided in Ref.~\cite{Note1}.

\begin{figure}[t]
\center
\includegraphics[width=8.6cm]{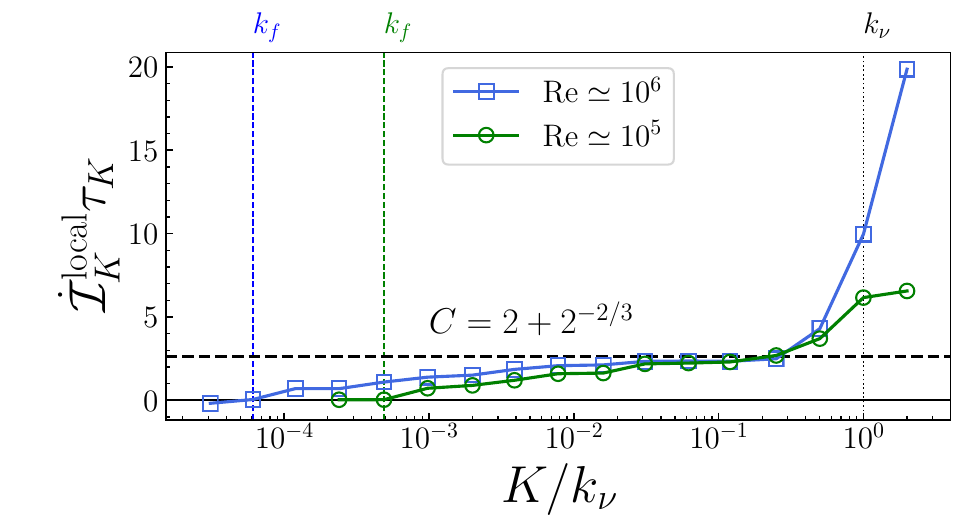}
\caption{Scale dependence of the information flow $\dot{\mathcal{I}}^{\mathrm{local}}_K(\simeq\dot{\mathcal{I}}_K)$ scaled by $\tau_K$.
The dashed horizontal line represents $C=2+2^{-2/3}$.
The vertical dotted and dashed lines represent the dissipation scale $k_\nu$ and injection scale $k_f$, respectively.
}
\label{fig:numerical demonstration}
\end{figure}

Figure \ref{fig:numerical demonstration} shows the scale dependence of the estimated $\dot{\mathcal{I}}^{\mathrm{local}}_K$ scaled by $\tau_K$.
Because it is difficult to estimate $C_p$ even in numerical simulations, we plot the value $C_p=2+2^{-2/3}=:C$ obtained under the additional assumption.
From this figure, we can see that Eq.~(\ref{IF upper bound}) is satisfied in the inertial range.
(Note that $\left\langle|u_{n_K}|^p\right\rangle^{1/p}\le\left\langle|u_{n_K}|^q\right\rangle^{1/q}$ for $p\le q$ from the H\"older inequality.)
Interestingly, the equality for $p=1$ is almost achieved, although it is based on the additional strong assumption.

Note that our estimated results are not entirely accurate.
First, the number of samples used in the KSG estimator, $N_{\mathrm{samp}}=3\times10^5$, is insufficient because the standard deviation of the estimated mutual information is estimated to be comparable to the finite difference $\Delta I$.
In other words, if we naively estimate the error of $\dot{\mathcal{I}}^{\mathrm{local}}_K$, it is of the same order as $\dot{\mathcal{I}}^{\mathrm{local}}_K$ itself.
In addition, the scale dependence of $\dot{\mathcal{I}}^{\mathrm{local}}_K$ at scales $\gtrsim k_\nu$ is not reliable because we cannot resolve the dynamics in this region with $\Delta t$ used in the estimation.
Consequently, the data in the far dissipation range are not shown in Fig.~\ref{fig:numerical demonstration}.

\paragraph*{Proof of the first main result (\ref{first main result}).}\parhyphen[0pt]
Here, we prove Eq.~(\ref{first main result}).
We first note that $\dot{I}^<_K$ can be expressed as~\cite{Note1}
\begin{align}
\dot{I}^<_K=\sum^{n_K}_{n=0}\left\langle\left[V_n(u,u^*)\dfrac{\partial}{\partial u_n}\ln\dfrac{p_t(u,u^*)}{p_t({\bm U}^<_K)p_t({\bm U}^>_K)}+\mathrm{c.c.}\right]\right\rangle,
% &=\sum^{n_K}_{n=0}\int dudu^*\left[J_n(u,u^*)\dfrac{\partial}{\partial u_n}\ln\dfrac{p_t(u,u^*)}{p_t({\bm U}^<_K)p_t({\bm U}^>_K)}+\mathrm{c.c.}\right],
\label{IF in terms of probability current}
\end{align}
where $V_n(u,u^*):=J_n(u,u^*)/p_t(u,u^*)$.
Because the viscous and thermal noise terms in $J_n(u,u^*)$ can be ignored in the inertial range and the contribution from $f_n$ vanishes~\cite{tanogami2024scale,Note1}, Eq.~(\ref{IF in terms of probability current}) can be rewritten as
\begin{align}
\dot{I}^<_K&=\sum^{n_K}_{n=0}\left\langle\left[B_n(u,u^*)\dfrac{\partial}{\partial u_n}\ln\dfrac{p_t(u,u^*)}{p_t({\bm U}^<_K)p_t({\bm U}^>_K)}+\mathrm{c.c.}\right]\right\rangle\notag\\
% &=\sum^{n_K}_{n=0}\int dudu^*\left[B_n(u,u^*)p_t(u,u^*)\right.\notag\\
% &\qquad\qquad\left.\times\dfrac{\partial}{\partial u_n}\ln\dfrac{p_t(u,u^*)}{p_t({\bm U}^<_K)p_t({\bm U}^>_K)}+\mathrm{c.c.}\right]\notag\\
&=\sum^{n_K}_{n=0}\int dudu^*\left[B_n(u,u^*)\dfrac{\partial}{\partial u_n}p_t(u,u^*)\right.\notag\\
&\left.\quad-B_n(u,u^*)p_t({\bm U}^>_K|{\bm U}^<_K)\dfrac{\partial}{\partial u_n}p_t({\bm U}^<_K)+\mathrm{c.c.}\right].
\end{align}
By integrating by parts and using Eqs.~(\ref{effective nonlinear term}) and (\ref{Liouville theorem}), we arrive at Eq.~(\ref{first main result}).

\paragraph*{Proof of the second main result (\ref{IF upper bound}).}\parhyphen[0pt]
The second main result (\ref{IF upper bound}) is derived from the first main result (\ref{first main result}).
We evaluate $\dot{\sigma}_K$ by using several properties of $B_n(u,u^*)$.
First, we note that because $B_n(u,u^*)$ is a homogeneous function of its arguments of degree two and possesses the phase symmetry~\cite{l1998improved}, $B_n(u,u^*)$ and $\overline{B}_n({\bm U}^<_K)$ for $0\le n\le n_K$ can be rewritten as
\begin{align}
B_n(u,u^*)&=k_n|u_n|^2e^{i\theta_n}\mathcal{B}_n(z,z^*),\\
\overline{B}_n({\bm U}^<_K)&=k_n|u_n|^2e^{i\theta_n}\overline{\mathcal{B}}_n({\bm Z}^{<,\mathrm{local}}_K),
\label{effective_B_nK}
\end{align}
where $\mathcal{B}_n(z,z^*)$ is a function of $\{z,z^*\}$, and $\overline{\mathcal{B}}_n({\bm Z}^{<,\mathrm{local}}_K)$ denotes its conditional average with respect to $p_{\mathrm{uni}}({\bm Z}^>_K|{\bm Z}^{<,\mathrm{local}}_K)$.
Here, we used Eq.~(\ref{Kolmogorov hypothesis}) by assuming Kolmogorov's hypothesis.
Note that, for $n\le n_K-2$, $\overline{B}_n({\bm U}^<_K)=B_n(u,u^*)$ because $B_n(u,u^*)$ does not depend on ${\bm U}^>_K$.

From Eqs.~(\ref{Liouville theorem}) and (\ref{effective_B_nK}), we can calculate the divergence of $\overline{B}_n({\bm U}^<_K)$ as
\begin{align}
&\dfrac{\partial}{\partial u_{n}}\overline{B}_{n}({\bm U}^<_K)+\dfrac{\partial}{\partial u^*_{n}}\overline{B}^*_{n}({\bm U}^<_K)\notag\\
&=
\begin{cases}
0\quad&\text{for}\quad n=0,1,\ldots,n_K-2,\\
K|u_{n_K}|g_n({\bm Z}^{<,\mathrm{local}}_K)\quad&\text{for}\quad n=n_K-1,n_K,
\end{cases}
\label{derivative of the effective nonlinear term}
\end{align}
where $g_n({\bm Z}^{<,\mathrm{local}}_K)$ is a universal real-valued function of ${\bm Z}^{<,\mathrm{local}}_K$.
See Ref.~\cite{Note1} for the specific form of $g_n({\bm Z}^{<,\mathrm{local}}_K)$ and the derivation of Eq.~(\ref{derivative of the effective nonlinear term}).
Substituting Eq.~(\ref{derivative of the effective nonlinear term}) into Eq.~(\ref{average phase-space contraction rate}), we obtain
\begin{align}
\dot{\mathcal{I}}_K=K\biggl\langle|u_{n_K}|G({\bm Z}^{<,\mathrm{local}}_K)\biggr\rangle,
% \quad \text{for}\quad k_f\ll K\ll k_\nu,
\label{IF equality}
\end{align}
where $G({\bm Z}^{<,\mathrm{local}}_K):=-g_{n_K-1}({\bm Z}^{<,\mathrm{local}}_K)-g_{n_K}({\bm Z}^{<,\mathrm{local}}_K)$.
By using the H\"older inequality, we arrive at Eq.~(\ref{IF upper bound}) with
\begin{align}
C_p:=\left\langle|G({\bm Z}^{<,\mathrm{local}}_K)|^q\right\rangle^{1/q}\quad\text{with}\quad\dfrac{1}{p}+\dfrac{1}{q}=1,
\label{def C_p}
\end{align}
which is a universal $K$-independent constant under Kolmogorov's assumption.

\paragraph*{Concluding remarks.}\parhyphen[0pt]
% summary
Here, we discuss the difference between the current and previous results.
In Ref.~\cite{tanogami2024information}, from the numerical simulation result, we have conjectured that $\dot{\mathcal{I}}_K\sim C'\tau^{-1}_L$ in the inertial range, where $C'$ is a universal constant. 
This scaling is inconsistent with the current analytical and numerical results.
This discrepancy may be due to a bias that exists in the previous numerical estimation, as already pointed out in Ref.~\cite{tanogami2024information}.
In the previous numerical estimation, we used the original definition of $\dot{\mathcal{I}}_K$, which includes multiscale shell variables, whereas in the current numerical estimation, we used $\dot{\mathcal{I}}^{\mathrm{local}}_K$ based on the scale locality~\cite{tanogami2024scale}.
Because $\dot{\mathcal{I}}^{\mathrm{local}}_K$ includes fewer shell variables than $\dot{\mathcal{I}}_K$, it can be easily estimated without significant bias~\cite{Note1}.

% remarks: extension
Although our results are derived only for the shell model, we conjecture that similar results can be proved for the Navier--Stokes equation because Kolmogorov's hypothesis is confirmed numerically and experimentally~\cite{chen2003kolmogorov,benzi1998multiscale,pedrizzetti1996self,chhabra1992scale}.
% It would also be an interesting research direction to investigate the relation between the information flow and the inverse error cascade, which causes an intrinsic finite range of predictability~\cite{lorenz1969predictability,palmer2014real,bandak2024spontaneous,palmer2024real}.
We also conjecture that the information flow is related to the generation mechanism of small-scale intermittency because the intermittency implies that turbulent fluctuations build up during the cascade process and ``remember'' the largest scale~\cite{Frisch,Eyink_Sreenivasan,Eyink_lecture}.
It would be an interesting research direction to investigate these points.

\begin{acknowledgments}
T.T.~thanks Takeshi Matsumoto for fruitful discussions, especially for pointing out the relevance of our results to Landau's objection.
T.T.~also thanks Shin-ichi Sasa for useful discussions on the KS entropy.
T.T.~was supported by JSPS KAKENHI Grant Number JP23K19035 and JST PRESTO Grant Number JPMJPR23O6, Japan.
R.A.~was supported by JSPS KAKENHI Grant Number 24K22942, Japan.
\end{acknowledgments}

\bibliography{main_text}

\begin{thebibliography}{75}%
\makeatletter
\providecommand \@ifxundefined [1]{%
 \@ifx{#1\undefined}
}%
\providecommand \@ifnum [1]{%
 \ifnum #1\expandafter \@firstoftwo
 \else \expandafter \@secondoftwo
 \fi
}%
\providecommand \@ifx [1]{%
 \ifx #1\expandafter \@firstoftwo
 \else \expandafter \@secondoftwo
 \fi
}%
\providecommand \natexlab [1]{#1}%
\providecommand \enquote  [1]{``#1''}%
\providecommand \bibnamefont  [1]{#1}%
\providecommand \bibfnamefont [1]{#1}%
\providecommand \citenamefont [1]{#1}%
\providecommand \href@noop [0]{\@secondoftwo}%
\providecommand \href [0]{\begingroup \@sanitize@url \@href}%
\providecommand \@href[1]{\@@startlink{#1}\@@href}%
\providecommand \@@href[1]{\endgroup#1\@@endlink}%
\providecommand \@sanitize@url [0]{\catcode `\\12\catcode `\$12\catcode
  `\&12\catcode `\#12\catcode `\^12\catcode `\_12\catcode `\%12\relax}%
\providecommand \@@startlink[1]{}%
\providecommand \@@endlink[0]{}%
\providecommand \url  [0]{\begingroup\@sanitize@url \@url }%
\providecommand \@url [1]{\endgroup\@href {#1}{\urlprefix }}%
\providecommand \urlprefix  [0]{URL }%
\providecommand \Eprint [0]{\href }%
\providecommand \doibase [0]{https://doi.org/}%
\providecommand \selectlanguage [0]{\@gobble}%
\providecommand \bibinfo  [0]{\@secondoftwo}%
\providecommand \bibfield  [0]{\@secondoftwo}%
\providecommand \translation [1]{[#1]}%
\providecommand \BibitemOpen [0]{}%
\providecommand \bibitemStop [0]{}%
\providecommand \bibitemNoStop [0]{.\EOS\space}%
\providecommand \EOS [0]{\spacefactor3000\relax}%
\providecommand \BibitemShut  [1]{\csname bibitem#1\endcsname}%
\let\auto@bib@innerbib\@empty
%</preamble>
\bibitem [{\citenamefont {Lorenz}(1969)}]{lorenz1969predictability}%
  \BibitemOpen
  \bibfield  {author} {\bibinfo {author} {\bibfnamefont {E.~N.}\ \bibnamefont
  {Lorenz}},\ }\bibfield  {title} {\bibinfo {title} {The predictability of a
  flow which possesses many scales of motion},\ }\href@noop {} {\bibfield
  {journal} {\bibinfo  {journal} {Tellus}\ }\textbf {\bibinfo {volume} {21}},\
  \bibinfo {pages} {289} (\bibinfo {year} {1969})}\BibitemShut {NoStop}%
\bibitem [{\citenamefont {Palmer}\ \emph {et~al.}(2014)\citenamefont {Palmer},
  \citenamefont {D{\"o}ring},\ and\ \citenamefont {Seregin}}]{palmer2014real}%
  \BibitemOpen
  \bibfield  {author} {\bibinfo {author} {\bibfnamefont {T.}~\bibnamefont
  {Palmer}}, \bibinfo {author} {\bibfnamefont {A.}~\bibnamefont {D{\"o}ring}},\
  and\ \bibinfo {author} {\bibfnamefont {G.}~\bibnamefont {Seregin}},\
  }\bibfield  {title} {\bibinfo {title} {The real butterfly effect},\
  }\href@noop {} {\bibfield  {journal} {\bibinfo  {journal} {Nonlinearity}\
  }\textbf {\bibinfo {volume} {27}},\ \bibinfo {pages} {R123} (\bibinfo {year}
  {2014})}\BibitemShut {NoStop}%
\bibitem [{\citenamefont {Palmer}(2024)}]{palmer2024real}%
  \BibitemOpen
  \bibfield  {author} {\bibinfo {author} {\bibfnamefont {T.}~\bibnamefont
  {Palmer}},\ }\bibfield  {title} {\bibinfo {title} {The real butterfly effect
  and maggoty apples},\ }\href@noop {} {\bibfield  {journal} {\bibinfo
  {journal} {Phys. Today}\ }\textbf {\bibinfo {volume} {77}},\ \bibinfo {pages}
  {30} (\bibinfo {year} {2024})}\BibitemShut {NoStop}%
\bibitem [{\citenamefont {Frisch}(1995)}]{Frisch}%
  \BibitemOpen
  \bibfield  {author} {\bibinfo {author} {\bibfnamefont {U.}~\bibnamefont
  {Frisch}},\ }\href@noop {} {\emph {\bibinfo {title} {Turbulence}}}\ (\bibinfo
   {publisher} {Cambridge university press},\ \bibinfo {year}
  {1995})\BibitemShut {NoStop}%
\bibitem [{\citenamefont {Bohr}\ \emph {et~al.}(1998)\citenamefont {Bohr},
  \citenamefont {Jensen}, \citenamefont {Paladin},\ and\ \citenamefont
  {Vulpiani}}]{bohr1998dynamical}%
  \BibitemOpen
  \bibfield  {author} {\bibinfo {author} {\bibfnamefont {T.}~\bibnamefont
  {Bohr}}, \bibinfo {author} {\bibfnamefont {M.~H.}\ \bibnamefont {Jensen}},
  \bibinfo {author} {\bibfnamefont {G.}~\bibnamefont {Paladin}},\ and\ \bibinfo
  {author} {\bibfnamefont {A.}~\bibnamefont {Vulpiani}},\ }\href@noop {} {\emph
  {\bibinfo {title} {Dynamical systems approach to turbulence}}}\ (\bibinfo
  {publisher} {Cambridge university press},\ \bibinfo {year}
  {1998})\BibitemShut {NoStop}%
\bibitem [{\citenamefont {Davidson}(2015)}]{davidson2015turbulence}%
  \BibitemOpen
  \bibfield  {author} {\bibinfo {author} {\bibfnamefont {P.~A.}\ \bibnamefont
  {Davidson}},\ }\href@noop {} {\emph {\bibinfo {title} {Turbulence: {A}n
  {I}ntroduction for {S}cientists and {E}ngineers}}},\ \bibinfo {edition}
  {2nd}\ ed.\ (\bibinfo  {publisher} {Oxford University Press},\ \bibinfo
  {year} {2015})\BibitemShut {NoStop}%
\bibitem [{\citenamefont {Eyink}()}]{Eyink_lecture}%
  \BibitemOpen
  \bibfield  {author} {\bibinfo {author} {\bibfnamefont {G.~L.}\ \bibnamefont
  {Eyink}},\ }\href@noop {} {\bibinfo {title} {Turbulence {T}heory, {C}ourse
  {N}otes}},\ \bibinfo {note}
  {\url{http://www.ams.jhu.edu/~eyink/Turbulence/notes/}}\BibitemShut {NoStop}%
\bibitem [{\citenamefont {Landau}\ and\ \citenamefont
  {Lifshitz}(1959)}]{landau1959fluid}%
  \BibitemOpen
  \bibfield  {author} {\bibinfo {author} {\bibfnamefont {L.~D.}\ \bibnamefont
  {Landau}}\ and\ \bibinfo {author} {\bibfnamefont {E.~M.}\ \bibnamefont
  {Lifshitz}},\ }\href@noop {} {\emph {\bibinfo {title} {Fluid {M}echanics}}},\
  Vol.~\bibinfo {volume} {6}\ (\bibinfo  {publisher} {Addision-Wesley, Reading,
  MA},\ \bibinfo {year} {1959})\BibitemShut {NoStop}%
\bibitem [{\citenamefont {Kraichnan}(1974)}]{kraichnan1974kolmogorov}%
  \BibitemOpen
  \bibfield  {author} {\bibinfo {author} {\bibfnamefont {R.~H.}\ \bibnamefont
  {Kraichnan}},\ }\bibfield  {title} {\bibinfo {title} {{On Kolmogorov's
  inertial-range theories}},\ }\href@noop {} {\bibfield  {journal} {\bibinfo
  {journal} {J. Fluid Mech.}\ }\textbf {\bibinfo {volume} {62}},\ \bibinfo
  {pages} {305} (\bibinfo {year} {1974})}\BibitemShut {NoStop}%
\bibitem [{\citenamefont {de~Wit}\ \emph {et~al.}(2024)\citenamefont {de~Wit},
  \citenamefont {Ortali}, \citenamefont {Corbetta}, \citenamefont {Mailybaev},
  \citenamefont {Biferale},\ and\ \citenamefont {Toschi}}]{de2024extreme}%
  \BibitemOpen
  \bibfield  {author} {\bibinfo {author} {\bibfnamefont {X.~M.}\ \bibnamefont
  {de~Wit}}, \bibinfo {author} {\bibfnamefont {G.}~\bibnamefont {Ortali}},
  \bibinfo {author} {\bibfnamefont {A.}~\bibnamefont {Corbetta}}, \bibinfo
  {author} {\bibfnamefont {A.~A.}\ \bibnamefont {Mailybaev}}, \bibinfo {author}
  {\bibfnamefont {L.}~\bibnamefont {Biferale}},\ and\ \bibinfo {author}
  {\bibfnamefont {F.}~\bibnamefont {Toschi}},\ }\bibfield  {title} {\bibinfo
  {title} {Extreme statistics and extreme events in dynamical models of
  turbulence},\ }\href@noop {} {\bibfield  {journal} {\bibinfo  {journal}
  {Phys. Rev. E}\ }\textbf {\bibinfo {volume} {109}},\ \bibinfo {pages}
  {055106} (\bibinfo {year} {2024})}\BibitemShut {NoStop}%
\bibitem [{\citenamefont {Cover}\ and\ \citenamefont
  {Thomas}(2006)}]{cover1999elements}%
  \BibitemOpen
  \bibfield  {author} {\bibinfo {author} {\bibfnamefont {T.~M.}\ \bibnamefont
  {Cover}}\ and\ \bibinfo {author} {\bibfnamefont {J.~A.}\ \bibnamefont
  {Thomas}},\ }\href@noop {} {\emph {\bibinfo {title} {Elements of
  {I}nformation {T}heory}}},\ \bibinfo {edition} {2nd}\ ed.\ (\bibinfo
  {publisher} {Wiley-Interscience, Hoboken, NJ},\ \bibinfo {year}
  {2006})\BibitemShut {NoStop}%
\bibitem [{\citenamefont {Peliti}\ and\ \citenamefont
  {Pigolotti}(2021)}]{peliti2021stochastic}%
  \BibitemOpen
  \bibfield  {author} {\bibinfo {author} {\bibfnamefont {L.}~\bibnamefont
  {Peliti}}\ and\ \bibinfo {author} {\bibfnamefont {S.}~\bibnamefont
  {Pigolotti}},\ }\href@noop {} {\emph {\bibinfo {title} {Stochastic
  {T}hermodynamics: {A}n {I}ntroduction}}}\ (\bibinfo  {publisher} {Princeton
  University Press},\ \bibinfo {year} {2021})\BibitemShut {NoStop}%
\bibitem [{\citenamefont {Shiraishi}(2023)}]{shiraishi2023introduction}%
  \BibitemOpen
  \bibfield  {author} {\bibinfo {author} {\bibfnamefont {N.}~\bibnamefont
  {Shiraishi}},\ }\href@noop {} {\emph {\bibinfo {title} {{An Introduction to
  Stochastic Thermodynamics: From Basic to Advanced}}}},\ Vol.\ \bibinfo
  {volume} {212}\ (\bibinfo  {publisher} {Springer Nature},\ \bibinfo {year}
  {2023})\BibitemShut {NoStop}%
\bibitem [{\citenamefont {Parrondo}\ \emph {et~al.}(2015)\citenamefont
  {Parrondo}, \citenamefont {Horowitz},\ and\ \citenamefont
  {Sagawa}}]{parrondo2015thermodynamics}%
  \BibitemOpen
  \bibfield  {author} {\bibinfo {author} {\bibfnamefont {J.~M.}\ \bibnamefont
  {Parrondo}}, \bibinfo {author} {\bibfnamefont {J.~M.}\ \bibnamefont
  {Horowitz}},\ and\ \bibinfo {author} {\bibfnamefont {T.}~\bibnamefont
  {Sagawa}},\ }\bibfield  {title} {\bibinfo {title} {Thermodynamics of
  information},\ }\href@noop {} {\bibfield  {journal} {\bibinfo  {journal}
  {Nat. Phys.}\ }\textbf {\bibinfo {volume} {11}},\ \bibinfo {pages} {131}
  (\bibinfo {year} {2015})}\BibitemShut {NoStop}%
\bibitem [{\citenamefont {Horowitz}\ and\ \citenamefont
  {Esposito}(2014)}]{horowitz2014thermodynamics}%
  \BibitemOpen
  \bibfield  {author} {\bibinfo {author} {\bibfnamefont {J.~M.}\ \bibnamefont
  {Horowitz}}\ and\ \bibinfo {author} {\bibfnamefont {M.}~\bibnamefont
  {Esposito}},\ }\bibfield  {title} {\bibinfo {title} {Thermodynamics with
  continuous information flow},\ }\href@noop {} {\bibfield  {journal} {\bibinfo
   {journal} {Phys. Rev. X}\ }\textbf {\bibinfo {volume} {4}},\ \bibinfo
  {pages} {031015} (\bibinfo {year} {2014})}\BibitemShut {NoStop}%
\bibitem [{\citenamefont {Horowitz}(2015)}]{horowitz2015multipartite}%
  \BibitemOpen
  \bibfield  {author} {\bibinfo {author} {\bibfnamefont {J.~M.}\ \bibnamefont
  {Horowitz}},\ }\bibfield  {title} {\bibinfo {title} {Multipartite information
  flow for multiple {M}axwell demons},\ }\href@noop {} {\bibfield  {journal}
  {\bibinfo  {journal} {J. Stat. Mech.}\ ,\ \bibinfo {pages} {P03006}}
  (\bibinfo {year} {2015})}\BibitemShut {NoStop}%
\bibitem [{\citenamefont {Ehrich}\ and\ \citenamefont
  {Sivak}(2023)}]{ehrich2023energy}%
  \BibitemOpen
  \bibfield  {author} {\bibinfo {author} {\bibfnamefont {J.}~\bibnamefont
  {Ehrich}}\ and\ \bibinfo {author} {\bibfnamefont {D.~A.}\ \bibnamefont
  {Sivak}},\ }\bibfield  {title} {\bibinfo {title} {Energy and information
  flows in autonomous systems},\ }\href@noop {} {\bibfield  {journal} {\bibinfo
   {journal} {Front. Phys.}\ }\textbf {\bibinfo {volume} {11}},\ \bibinfo
  {pages} {155} (\bibinfo {year} {2023})}\BibitemShut {NoStop}%
\bibitem [{\citenamefont {Tanogami}\ \emph {et~al.}(2023)\citenamefont
  {Tanogami}, \citenamefont {Van~Vu},\ and\ \citenamefont
  {Saito}}]{tanogami2023universal}%
  \BibitemOpen
  \bibfield  {author} {\bibinfo {author} {\bibfnamefont {T.}~\bibnamefont
  {Tanogami}}, \bibinfo {author} {\bibfnamefont {T.}~\bibnamefont {Van~Vu}},\
  and\ \bibinfo {author} {\bibfnamefont {K.}~\bibnamefont {Saito}},\ }\bibfield
   {title} {\bibinfo {title} {Universal bounds on the performance of
  information-thermodynamic engine},\ }\href@noop {} {\bibfield  {journal}
  {\bibinfo  {journal} {Phys. Rev. Research}\ }\textbf {\bibinfo {volume}
  {5}},\ \bibinfo {pages} {043280} (\bibinfo {year} {2023})}\BibitemShut
  {NoStop}%
\bibitem [{\citenamefont {Goold}\ \emph {et~al.}(2016)\citenamefont {Goold},
  \citenamefont {Huber}, \citenamefont {Riera}, \citenamefont {Del~Rio},\ and\
  \citenamefont {Skrzypczyk}}]{goold2016role}%
  \BibitemOpen
  \bibfield  {author} {\bibinfo {author} {\bibfnamefont {J.}~\bibnamefont
  {Goold}}, \bibinfo {author} {\bibfnamefont {M.}~\bibnamefont {Huber}},
  \bibinfo {author} {\bibfnamefont {A.}~\bibnamefont {Riera}}, \bibinfo
  {author} {\bibfnamefont {L.}~\bibnamefont {Del~Rio}},\ and\ \bibinfo {author}
  {\bibfnamefont {P.}~\bibnamefont {Skrzypczyk}},\ }\bibfield  {title}
  {\bibinfo {title} {The role of quantum information in thermodynamics―a
  topical review},\ }\href@noop {} {\bibfield  {journal} {\bibinfo  {journal}
  {J. Phys. A}\ }\textbf {\bibinfo {volume} {49}},\ \bibinfo {pages} {143001}
  (\bibinfo {year} {2016})}\BibitemShut {NoStop}%
\bibitem [{\citenamefont {Gong}\ and\ \citenamefont
  {Hamazaki}(2022)}]{gong2022bounds}%
  \BibitemOpen
  \bibfield  {author} {\bibinfo {author} {\bibfnamefont {Z.}~\bibnamefont
  {Gong}}\ and\ \bibinfo {author} {\bibfnamefont {R.}~\bibnamefont
  {Hamazaki}},\ }\bibfield  {title} {\bibinfo {title} {Bounds in nonequilibrium
  quantum dynamics},\ }\href@noop {} {\bibfield  {journal} {\bibinfo  {journal}
  {Int. J. Mod. Phys. B}\ }\textbf {\bibinfo {volume} {36}},\ \bibinfo {pages}
  {2230007} (\bibinfo {year} {2022})}\BibitemShut {NoStop}%
\bibitem [{\citenamefont {Betchov}(1964)}]{betchov1964measure}%
  \BibitemOpen
  \bibfield  {author} {\bibinfo {author} {\bibfnamefont {R.}~\bibnamefont
  {Betchov}},\ }\bibfield  {title} {\bibinfo {title} {Measure of the intricacy
  of turbulence},\ }\href@noop {} {\bibfield  {journal} {\bibinfo  {journal}
  {Phys. Fluids}\ }\textbf {\bibinfo {volume} {7}},\ \bibinfo {pages} {1160}
  (\bibinfo {year} {1964})}\BibitemShut {NoStop}%
\bibitem [{\citenamefont {Ikeda}\ and\ \citenamefont
  {Matsumoto}(1989)}]{ikeda1989information}%
  \BibitemOpen
  \bibfield  {author} {\bibinfo {author} {\bibfnamefont {K.}~\bibnamefont
  {Ikeda}}\ and\ \bibinfo {author} {\bibfnamefont {K.}~\bibnamefont
  {Matsumoto}},\ }\bibfield  {title} {\bibinfo {title} {Information theoretical
  characterization of turbulence},\ }\href@noop {} {\bibfield  {journal}
  {\bibinfo  {journal} {Phys. Rev. Lett.}\ }\textbf {\bibinfo {volume} {62}},\
  \bibinfo {pages} {2265} (\bibinfo {year} {1989})}\BibitemShut {NoStop}%
\bibitem [{\citenamefont {Cerbus}\ and\ \citenamefont
  {Goldburg}(2013)}]{cerbus2013information}%
  \BibitemOpen
  \bibfield  {author} {\bibinfo {author} {\bibfnamefont {R.~T.}\ \bibnamefont
  {Cerbus}}\ and\ \bibinfo {author} {\bibfnamefont {W.~I.}\ \bibnamefont
  {Goldburg}},\ }\bibfield  {title} {\bibinfo {title} {Information content of
  turbulence},\ }\href@noop {} {\bibfield  {journal} {\bibinfo  {journal}
  {Phys. Rev. E}\ }\textbf {\bibinfo {volume} {88}},\ \bibinfo {pages} {053012}
  (\bibinfo {year} {2013})}\BibitemShut {NoStop}%
\bibitem [{\citenamefont {Materassi}\ \emph {et~al.}(2014)\citenamefont
  {Materassi}, \citenamefont {Consolini}, \citenamefont {Smith},\ and\
  \citenamefont {De~Marco}}]{materassi2014information}%
  \BibitemOpen
  \bibfield  {author} {\bibinfo {author} {\bibfnamefont {M.}~\bibnamefont
  {Materassi}}, \bibinfo {author} {\bibfnamefont {G.}~\bibnamefont
  {Consolini}}, \bibinfo {author} {\bibfnamefont {N.}~\bibnamefont {Smith}},\
  and\ \bibinfo {author} {\bibfnamefont {R.}~\bibnamefont {De~Marco}},\
  }\bibfield  {title} {\bibinfo {title} {Information theory analysis of
  cascading process in a synthetic model of fluid turbulence},\ }\href@noop {}
  {\bibfield  {journal} {\bibinfo  {journal} {Entropy}\ }\textbf {\bibinfo
  {volume} {16}},\ \bibinfo {pages} {1272} (\bibinfo {year}
  {2014})}\BibitemShut {NoStop}%
\bibitem [{\citenamefont {Cerbus}\ and\ \citenamefont
  {Goldburg}(2016)}]{cerbus2016information}%
  \BibitemOpen
  \bibfield  {author} {\bibinfo {author} {\bibfnamefont {R.~T.}\ \bibnamefont
  {Cerbus}}\ and\ \bibinfo {author} {\bibfnamefont {W.~I.}\ \bibnamefont
  {Goldburg}},\ }\bibfield  {title} {\bibinfo {title} {Information theory
  demonstration of the {R}ichardson cascade},\ }\href@noop {} {\bibfield
  {journal} {\bibinfo  {journal} {arXiv preprint arXiv:1602.02980}\ } (\bibinfo
  {year} {2016})}\BibitemShut {NoStop}%
\bibitem [{\citenamefont {Goldburg}\ and\ \citenamefont
  {Cerbus}(2016)}]{goldburg2016turbulence}%
  \BibitemOpen
  \bibfield  {author} {\bibinfo {author} {\bibfnamefont {W.~I.}\ \bibnamefont
  {Goldburg}}\ and\ \bibinfo {author} {\bibfnamefont {R.~T.}\ \bibnamefont
  {Cerbus}},\ }\bibfield  {title} {\bibinfo {title} {Turbulence as
  information},\ }\href@noop {} {\bibfield  {journal} {\bibinfo  {journal}
  {arXiv preprint arXiv:1609.00471}\ } (\bibinfo {year} {2016})}\BibitemShut
  {NoStop}%
\bibitem [{\citenamefont {Granero-Belinchon}\ \emph {et~al.}(2016)\citenamefont
  {Granero-Belinchon}, \citenamefont {Roux},\ and\ \citenamefont
  {Garnier}}]{granero2016scaling}%
  \BibitemOpen
  \bibfield  {author} {\bibinfo {author} {\bibfnamefont {C.}~\bibnamefont
  {Granero-Belinchon}}, \bibinfo {author} {\bibfnamefont {S.~G.}\ \bibnamefont
  {Roux}},\ and\ \bibinfo {author} {\bibfnamefont {N.~B.}\ \bibnamefont
  {Garnier}},\ }\bibfield  {title} {\bibinfo {title} {Scaling of information in
  turbulence},\ }\href@noop {} {\bibfield  {journal} {\bibinfo  {journal}
  {Europhys. Lett.}\ }\textbf {\bibinfo {volume} {115}},\ \bibinfo {pages}
  {58003} (\bibinfo {year} {2016})}\BibitemShut {NoStop}%
\bibitem [{\citenamefont {Granero-Belinch{\'o}n}\ \emph
  {et~al.}(2018)\citenamefont {Granero-Belinch{\'o}n}, \citenamefont {Roux},\
  and\ \citenamefont {Garnier}}]{granero2018kullback}%
  \BibitemOpen
  \bibfield  {author} {\bibinfo {author} {\bibfnamefont {C.}~\bibnamefont
  {Granero-Belinch{\'o}n}}, \bibinfo {author} {\bibfnamefont {S.~G.}\
  \bibnamefont {Roux}},\ and\ \bibinfo {author} {\bibfnamefont {N.~B.}\
  \bibnamefont {Garnier}},\ }\bibfield  {title} {\bibinfo {title}
  {Kullback-{L}eibler divergence measure of intermittency: {A}pplication to
  turbulence},\ }\href@noop {} {\bibfield  {journal} {\bibinfo  {journal}
  {Phys. Rev. E}\ }\textbf {\bibinfo {volume} {97}},\ \bibinfo {pages} {013107}
  (\bibinfo {year} {2018})}\BibitemShut {NoStop}%
\bibitem [{\citenamefont {Lozano-Dur{\'a}n}\ \emph {et~al.}(2020)\citenamefont
  {Lozano-Dur{\'a}n}, \citenamefont {Bae},\ and\ \citenamefont
  {Encinar}}]{lozano2020causality}%
  \BibitemOpen
  \bibfield  {author} {\bibinfo {author} {\bibfnamefont {A.}~\bibnamefont
  {Lozano-Dur{\'a}n}}, \bibinfo {author} {\bibfnamefont {H.~J.}\ \bibnamefont
  {Bae}},\ and\ \bibinfo {author} {\bibfnamefont {M.~P.}\ \bibnamefont
  {Encinar}},\ }\bibfield  {title} {\bibinfo {title} {Causality of
  energy-containing eddies in wall turbulence},\ }\href@noop {} {\bibfield
  {journal} {\bibinfo  {journal} {J. Fluid Mech.}\ }\textbf {\bibinfo {volume}
  {882}} (\bibinfo {year} {2020})}\BibitemShut {NoStop}%
\bibitem [{\citenamefont {Shavit}\ and\ \citenamefont
  {Falkovich}(2020)}]{shavit2020singular}%
  \BibitemOpen
  \bibfield  {author} {\bibinfo {author} {\bibfnamefont {M.}~\bibnamefont
  {Shavit}}\ and\ \bibinfo {author} {\bibfnamefont {G.}~\bibnamefont
  {Falkovich}},\ }\bibfield  {title} {\bibinfo {title} {Singular measures and
  information capacity of turbulent cascades},\ }\href@noop {} {\bibfield
  {journal} {\bibinfo  {journal} {Phys. Rev. Lett.}\ }\textbf {\bibinfo
  {volume} {125}},\ \bibinfo {pages} {104501} (\bibinfo {year}
  {2020})}\BibitemShut {NoStop}%
\bibitem [{\citenamefont {Vladimirova}\ \emph {et~al.}(2021)\citenamefont
  {Vladimirova}, \citenamefont {Shavit},\ and\ \citenamefont
  {Falkovich}}]{vladimirova2021fibonacci}%
  \BibitemOpen
  \bibfield  {author} {\bibinfo {author} {\bibfnamefont {N.}~\bibnamefont
  {Vladimirova}}, \bibinfo {author} {\bibfnamefont {M.}~\bibnamefont
  {Shavit}},\ and\ \bibinfo {author} {\bibfnamefont {G.}~\bibnamefont
  {Falkovich}},\ }\bibfield  {title} {\bibinfo {title} {Fibonacci turbulence},\
  }\href@noop {} {\bibfield  {journal} {\bibinfo  {journal} {Phys. Rev. X}\
  }\textbf {\bibinfo {volume} {11}},\ \bibinfo {pages} {021063} (\bibinfo
  {year} {2021})}\BibitemShut {NoStop}%
\bibitem [{\citenamefont {Lozano-Dur\'an}\ and\ \citenamefont
  {Arranz}(2022)}]{PhysRevResearch.4.023195}%
  \BibitemOpen
  \bibfield  {author} {\bibinfo {author} {\bibfnamefont {A.}~\bibnamefont
  {Lozano-Dur\'an}}\ and\ \bibinfo {author} {\bibfnamefont {G.}~\bibnamefont
  {Arranz}},\ }\bibfield  {title} {\bibinfo {title} {Information-theoretic
  formulation of dynamical systems: {C}ausality, modeling, and control},\
  }\href {https://doi.org/10.1103/PhysRevResearch.4.023195} {\bibfield
  {journal} {\bibinfo  {journal} {Phys. Rev. Research}\ }\textbf {\bibinfo
  {volume} {4}},\ \bibinfo {pages} {023195} (\bibinfo {year}
  {2022})}\BibitemShut {NoStop}%
\bibitem [{\citenamefont {Arranz}\ and\ \citenamefont
  {Lozano-Dur{\'a}n}(2024)}]{arranz2024informative}%
  \BibitemOpen
  \bibfield  {author} {\bibinfo {author} {\bibfnamefont {G.}~\bibnamefont
  {Arranz}}\ and\ \bibinfo {author} {\bibfnamefont {A.}~\bibnamefont
  {Lozano-Dur{\'a}n}},\ }\bibfield  {title} {\bibinfo {title} {Informative and
  non-informative decomposition of turbulent flow fields},\ }\href@noop {}
  {\bibfield  {journal} {\bibinfo  {journal} {arXiv preprint arXiv:2402.11448}\
  } (\bibinfo {year} {2024})}\BibitemShut {NoStop}%
\bibitem [{\citenamefont {Araki}\ \emph {et~al.}(2024)\citenamefont {Araki},
  \citenamefont {Vela-Mart{\'\i}n},\ and\ \citenamefont
  {Lozano-Dur{\'a}n}}]{araki2024forgetfulness}%
  \BibitemOpen
  \bibfield  {author} {\bibinfo {author} {\bibfnamefont {R.}~\bibnamefont
  {Araki}}, \bibinfo {author} {\bibfnamefont {A.}~\bibnamefont
  {Vela-Mart{\'\i}n}},\ and\ \bibinfo {author} {\bibfnamefont {A.}~\bibnamefont
  {Lozano-Dur{\'a}n}},\ }\bibfield  {title} {\bibinfo {title} {Forgetfulness of
  turbulent energy cascade associated with different mechanisms},\ }in\
  \href@noop {} {\emph {\bibinfo {booktitle} {J. Phys. Conf.}}},\ Vol.\
  \bibinfo {volume} {2753}\ (\bibinfo {organization} {IOP Publishing},\
  \bibinfo {year} {2024})\ p.\ \bibinfo {pages} {012001}\BibitemShut {NoStop}%
\bibitem [{\citenamefont {Yatomi}\ and\ \citenamefont
  {Nakata}(2024)}]{yatomi2024quantum}%
  \BibitemOpen
  \bibfield  {author} {\bibinfo {author} {\bibfnamefont {G.}~\bibnamefont
  {Yatomi}}\ and\ \bibinfo {author} {\bibfnamefont {M.}~\bibnamefont
  {Nakata}},\ }\bibfield  {title} {\bibinfo {title} {Quantum-inspired
  information entropy in multi-field turbulence},\ }\href@noop {} {\bibfield
  {journal} {\bibinfo  {journal} {arXiv preprint arXiv:2407.09098}\ } (\bibinfo
  {year} {2024})}\BibitemShut {NoStop}%
\bibitem [{\citenamefont {Tanogami}\ and\ \citenamefont
  {Araki}(2024)}]{tanogami2024information}%
  \BibitemOpen
  \bibfield  {author} {\bibinfo {author} {\bibfnamefont {T.}~\bibnamefont
  {Tanogami}}\ and\ \bibinfo {author} {\bibfnamefont {R.}~\bibnamefont
  {Araki}},\ }\bibfield  {title} {\bibinfo {title} {Information-thermodynamic
  bound on information flow in turbulent cascade},\ }\href@noop {} {\bibfield
  {journal} {\bibinfo  {journal} {Phys. Rev. Research}\ }\textbf {\bibinfo
  {volume} {6}},\ \bibinfo {pages} {013090} (\bibinfo {year}
  {2024})}\BibitemShut {NoStop}%
\bibitem [{\citenamefont {Evans}\ and\ \citenamefont
  {Searles}(2002)}]{evans2002fluctuation}%
  \BibitemOpen
  \bibfield  {author} {\bibinfo {author} {\bibfnamefont {D.~J.}\ \bibnamefont
  {Evans}}\ and\ \bibinfo {author} {\bibfnamefont {D.~J.}\ \bibnamefont
  {Searles}},\ }\bibfield  {title} {\bibinfo {title} {The fluctuation
  theorem},\ }\href@noop {} {\bibfield  {journal} {\bibinfo  {journal} {Adv.
  Phys.}\ }\textbf {\bibinfo {volume} {51}},\ \bibinfo {pages} {1529} (\bibinfo
  {year} {2002})}\BibitemShut {NoStop}%
\bibitem [{\citenamefont {Rondoni}\ and\ \citenamefont
  {Mej{\'\i}a-Monasterio}(2007)}]{rondoni2007fluctuations}%
  \BibitemOpen
  \bibfield  {author} {\bibinfo {author} {\bibfnamefont {L.}~\bibnamefont
  {Rondoni}}\ and\ \bibinfo {author} {\bibfnamefont {C.}~\bibnamefont
  {Mej{\'\i}a-Monasterio}},\ }\bibfield  {title} {\bibinfo {title}
  {Fluctuations in nonequilibrium statistical mechanics: models, mathematical
  theory, physical mechanisms},\ }\href@noop {} {\bibfield  {journal} {\bibinfo
   {journal} {Nonlinearity}\ }\textbf {\bibinfo {volume} {20}},\ \bibinfo
  {pages} {R1} (\bibinfo {year} {2007})}\BibitemShut {NoStop}%
\bibitem [{\citenamefont {Baiesi}\ and\ \citenamefont
  {Falasco}(2015)}]{baiesi2015inflow}%
  \BibitemOpen
  \bibfield  {author} {\bibinfo {author} {\bibfnamefont {M.}~\bibnamefont
  {Baiesi}}\ and\ \bibinfo {author} {\bibfnamefont {G.}~\bibnamefont
  {Falasco}},\ }\bibfield  {title} {\bibinfo {title} {Inflow rate, a
  time-symmetric observable obeying fluctuation relations},\ }\href@noop {}
  {\bibfield  {journal} {\bibinfo  {journal} {Phys. Rev. E}\ }\textbf {\bibinfo
  {volume} {92}},\ \bibinfo {pages} {042162} (\bibinfo {year}
  {2015})}\BibitemShut {NoStop}%
\bibitem [{\citenamefont {Gallavotti}(2020)}]{gallavotti2020nonequilibrium}%
  \BibitemOpen
  \bibfield  {author} {\bibinfo {author} {\bibfnamefont {G.}~\bibnamefont
  {Gallavotti}},\ }\bibfield  {title} {\bibinfo {title} {Nonequilibrium and
  fluctuation relation},\ }\href@noop {} {\bibfield  {journal} {\bibinfo
  {journal} {J. Stat. Phys.}\ }\textbf {\bibinfo {volume} {180}},\ \bibinfo
  {pages} {172} (\bibinfo {year} {2020})}\BibitemShut {NoStop}%
\bibitem [{\citenamefont {Eckmann}\ and\ \citenamefont
  {Ruelle}(1985)}]{eckmann1985ergodic}%
  \BibitemOpen
  \bibfield  {author} {\bibinfo {author} {\bibfnamefont {J.-P.}\ \bibnamefont
  {Eckmann}}\ and\ \bibinfo {author} {\bibfnamefont {D.}~\bibnamefont
  {Ruelle}},\ }\bibfield  {title} {\bibinfo {title} {Ergodic theory of chaos
  and strange attractors},\ }\href@noop {} {\bibfield  {journal} {\bibinfo
  {journal} {Rev. Mod. Phys.}\ }\textbf {\bibinfo {volume} {57}},\ \bibinfo
  {pages} {617} (\bibinfo {year} {1985})}\BibitemShut {NoStop}%
\bibitem [{\citenamefont {Boffetta}\ \emph {et~al.}(2002)\citenamefont
  {Boffetta}, \citenamefont {Cencini}, \citenamefont {Falcioni},\ and\
  \citenamefont {Vulpiani}}]{boffetta2002predictability}%
  \BibitemOpen
  \bibfield  {author} {\bibinfo {author} {\bibfnamefont {G.}~\bibnamefont
  {Boffetta}}, \bibinfo {author} {\bibfnamefont {M.}~\bibnamefont {Cencini}},
  \bibinfo {author} {\bibfnamefont {M.}~\bibnamefont {Falcioni}},\ and\
  \bibinfo {author} {\bibfnamefont {A.}~\bibnamefont {Vulpiani}},\ }\bibfield
  {title} {\bibinfo {title} {Predictability: a way to characterize
  complexity},\ }\href@noop {} {\bibfield  {journal} {\bibinfo  {journal}
  {Phys. Rep.}\ }\textbf {\bibinfo {volume} {356}},\ \bibinfo {pages} {367}
  (\bibinfo {year} {2002})}\BibitemShut {NoStop}%
\bibitem [{\citenamefont {Ruelle}(1982)}]{ruelle1982large}%
  \BibitemOpen
  \bibfield  {author} {\bibinfo {author} {\bibfnamefont {D.}~\bibnamefont
  {Ruelle}},\ }\bibfield  {title} {\bibinfo {title} {Large volume limit of the
  distribution of characteristic exponents in turbulence},\ }\href@noop {}
  {\bibfield  {journal} {\bibinfo  {journal} {Commun. Math. Phys.}\ }\textbf
  {\bibinfo {volume} {87}},\ \bibinfo {pages} {287} (\bibinfo {year}
  {1982})}\BibitemShut {NoStop}%
\bibitem [{\citenamefont {Ruelle}(1984)}]{ruelle1984characteristic}%
  \BibitemOpen
  \bibfield  {author} {\bibinfo {author} {\bibfnamefont {D.}~\bibnamefont
  {Ruelle}},\ }\bibfield  {title} {\bibinfo {title} {Characteristic exponents
  for a viscous fluid subjected to time dependent forces},\ }\href@noop {}
  {\bibfield  {journal} {\bibinfo  {journal} {Commun. Math. Phys.}\ }\textbf
  {\bibinfo {volume} {93}},\ \bibinfo {pages} {285} (\bibinfo {year}
  {1984})}\BibitemShut {NoStop}%
\bibitem [{\citenamefont {Berera}\ and\ \citenamefont
  {Clark}(2019)}]{berera2019information}%
  \BibitemOpen
  \bibfield  {author} {\bibinfo {author} {\bibfnamefont {A.}~\bibnamefont
  {Berera}}\ and\ \bibinfo {author} {\bibfnamefont {D.}~\bibnamefont {Clark}},\
  }\bibfield  {title} {\bibinfo {title} {Information production in homogeneous
  isotropic turbulence},\ }\href@noop {} {\bibfield  {journal} {\bibinfo
  {journal} {Phys. Rev. E}\ }\textbf {\bibinfo {volume} {100}},\ \bibinfo
  {pages} {041101} (\bibinfo {year} {2019})}\BibitemShut {NoStop}%
\bibitem [{\citenamefont {L'vov}\ \emph {et~al.}(1998)\citenamefont {L'vov},
  \citenamefont {Podivilov}, \citenamefont {Pomyalov}, \citenamefont
  {Procaccia},\ and\ \citenamefont {Vandembroucq}}]{l1998improved}%
  \BibitemOpen
  \bibfield  {author} {\bibinfo {author} {\bibfnamefont {V.~S.}\ \bibnamefont
  {L'vov}}, \bibinfo {author} {\bibfnamefont {E.}~\bibnamefont {Podivilov}},
  \bibinfo {author} {\bibfnamefont {A.}~\bibnamefont {Pomyalov}}, \bibinfo
  {author} {\bibfnamefont {I.}~\bibnamefont {Procaccia}},\ and\ \bibinfo
  {author} {\bibfnamefont {D.}~\bibnamefont {Vandembroucq}},\ }\bibfield
  {title} {\bibinfo {title} {Improved shell model of turbulence},\ }\href@noop
  {} {\bibfield  {journal} {\bibinfo  {journal} {Phys. Rev. E}\ }\textbf
  {\bibinfo {volume} {58}},\ \bibinfo {pages} {1811} (\bibinfo {year}
  {1998})}\BibitemShut {NoStop}%
\bibitem [{\citenamefont {Bandak}\ \emph {et~al.}(2021)\citenamefont {Bandak},
  \citenamefont {Eyink}, \citenamefont {Mailybaev},\ and\ \citenamefont
  {Goldenfeld}}]{bandak2021thermal}%
  \BibitemOpen
  \bibfield  {author} {\bibinfo {author} {\bibfnamefont {D.}~\bibnamefont
  {Bandak}}, \bibinfo {author} {\bibfnamefont {G.~L.}\ \bibnamefont {Eyink}},
  \bibinfo {author} {\bibfnamefont {A.}~\bibnamefont {Mailybaev}},\ and\
  \bibinfo {author} {\bibfnamefont {N.}~\bibnamefont {Goldenfeld}},\ }\bibfield
   {title} {\bibinfo {title} {Thermal noise competes with turbulent
  fluctuations below millimeter scales},\ }\href@noop {} {\bibfield  {journal}
  {\bibinfo  {journal} {arXiv preprint arXiv:2107.03184}\ } (\bibinfo {year}
  {2021})}\BibitemShut {NoStop}%
\bibitem [{\citenamefont {Bandak}\ \emph {et~al.}(2022)\citenamefont {Bandak},
  \citenamefont {Goldenfeld}, \citenamefont {Mailybaev},\ and\ \citenamefont
  {Eyink}}]{bandak2022dissipation}%
  \BibitemOpen
  \bibfield  {author} {\bibinfo {author} {\bibfnamefont {D.}~\bibnamefont
  {Bandak}}, \bibinfo {author} {\bibfnamefont {N.}~\bibnamefont {Goldenfeld}},
  \bibinfo {author} {\bibfnamefont {A.~A.}\ \bibnamefont {Mailybaev}},\ and\
  \bibinfo {author} {\bibfnamefont {G.}~\bibnamefont {Eyink}},\ }\bibfield
  {title} {\bibinfo {title} {Dissipation-range fluid turbulence and thermal
  noise},\ }\href@noop {} {\bibfield  {journal} {\bibinfo  {journal} {Phys.
  Rev. E}\ }\textbf {\bibinfo {volume} {105}},\ \bibinfo {pages} {065113}
  (\bibinfo {year} {2022})}\BibitemShut {NoStop}%
\bibitem [{\citenamefont {De~Zarate}\ and\ \citenamefont
  {Sengers}(2006)}]{de2006hydrodynamic}%
  \BibitemOpen
  \bibfield  {author} {\bibinfo {author} {\bibfnamefont {J.~M.~O.}\
  \bibnamefont {De~Zarate}}\ and\ \bibinfo {author} {\bibfnamefont {J.~V.}\
  \bibnamefont {Sengers}},\ }\href@noop {} {\emph {\bibinfo {title}
  {Hydrodynamic fluctuations in fluids and fluid mixtures}}}\ (\bibinfo
  {publisher} {Elsevier},\ \bibinfo {year} {2006})\BibitemShut {NoStop}%
\bibitem [{\citenamefont {Tanogami}(2024)}]{tanogami2024scale}%
  \BibitemOpen
  \bibfield  {author} {\bibinfo {author} {\bibfnamefont {T.}~\bibnamefont
  {Tanogami}},\ }\bibfield  {title} {\bibinfo {title} {Scale locality of
  information flow in turbulence},\ }\href@noop {} {\bibfield  {journal}
  {\bibinfo  {journal} {arXiv preprint arXiv:2407.20572}\ } (\bibinfo {year}
  {2024})}\BibitemShut {NoStop}%
\bibitem [{\citenamefont {Maes}(2021)}]{maes2021local}%
  \BibitemOpen
  \bibfield  {author} {\bibinfo {author} {\bibfnamefont {C.}~\bibnamefont
  {Maes}},\ }\bibfield  {title} {\bibinfo {title} {Local detailed balance},\
  }\href@noop {} {\bibfield  {journal} {\bibinfo  {journal} {SciPost Phys.
  Lect. Notes}\ }\textbf {\bibinfo {volume} {32}},\ \bibinfo {pages} {1}
  (\bibinfo {year} {2021})}\BibitemShut {NoStop}%
\bibitem [{\citenamefont {Risken}(1996)}]{risken1996fokker}%
  \BibitemOpen
  \bibfield  {author} {\bibinfo {author} {\bibfnamefont {H.}~\bibnamefont
  {Risken}},\ }\bibfield  {title} {\bibinfo {title} {The {F}okker-{P}lanck
  {E}quation}\ }(\bibinfo  {publisher} {Springer},\ \bibinfo {year}
  {1996})\BibitemShut {NoStop}%
\bibitem [{\citenamefont {Gardiner}(2009)}]{gardiner1985handbook}%
  \BibitemOpen
  \bibfield  {author} {\bibinfo {author} {\bibfnamefont {C.~W.}\ \bibnamefont
  {Gardiner}},\ }\href@noop {} {\emph {\bibinfo {title} {Handbook of
  {S}tochastic {M}ethods}}},\ \bibinfo {edition} {4th}\ ed.\ (\bibinfo
  {publisher} {Springer, Berlin},\ \bibinfo {year} {2009})\BibitemShut
  {NoStop}%
\bibitem [{\citenamefont {Biferale}(2003)}]{biferale2003shell}%
  \BibitemOpen
  \bibfield  {author} {\bibinfo {author} {\bibfnamefont {L.}~\bibnamefont
  {Biferale}},\ }\bibfield  {title} {\bibinfo {title} {Shell models of energy
  cascade in turbulence},\ }\href@noop {} {\bibfield  {journal} {\bibinfo
  {journal} {Annu. Rev. Fluid Mech.}\ }\textbf {\bibinfo {volume} {35}},\
  \bibinfo {pages} {441} (\bibinfo {year} {2003})}\BibitemShut {NoStop}%
\bibitem [{\citenamefont {Allahverdyan}\ \emph {et~al.}(2009)\citenamefont
  {Allahverdyan}, \citenamefont {Janzing},\ and\ \citenamefont
  {Mahler}}]{allahverdyan2009thermodynamic}%
  \BibitemOpen
  \bibfield  {author} {\bibinfo {author} {\bibfnamefont {A.~E.}\ \bibnamefont
  {Allahverdyan}}, \bibinfo {author} {\bibfnamefont {D.}~\bibnamefont
  {Janzing}},\ and\ \bibinfo {author} {\bibfnamefont {G.}~\bibnamefont
  {Mahler}},\ }\bibfield  {title} {\bibinfo {title} {Thermodynamic efficiency
  of information and heat flow},\ }\href@noop {} {\bibfield  {journal}
  {\bibinfo  {journal} {J. Stat. Mech.}\ }\textbf {\bibinfo {volume} {2009}},\
  \bibinfo {pages} {P09011} (\bibinfo {year} {2009})}\BibitemShut {NoStop}%
\bibitem [{\citenamefont {Hartich}\ \emph {et~al.}(2014)\citenamefont
  {Hartich}, \citenamefont {Barato},\ and\ \citenamefont
  {Seifert}}]{hartich2014stochastic}%
  \BibitemOpen
  \bibfield  {author} {\bibinfo {author} {\bibfnamefont {D.}~\bibnamefont
  {Hartich}}, \bibinfo {author} {\bibfnamefont {A.~C.}\ \bibnamefont
  {Barato}},\ and\ \bibinfo {author} {\bibfnamefont {U.}~\bibnamefont
  {Seifert}},\ }\bibfield  {title} {\bibinfo {title} {Stochastic thermodynamics
  of bipartite systems: transfer entropy inequalities and a {M}axwell’s demon
  interpretation},\ }\href@noop {} {\bibfield  {journal} {\bibinfo  {journal}
  {J. Stat. Mech.}\ }\textbf {\bibinfo {volume} {2014}},\ \bibinfo {pages}
  {P02016} (\bibinfo {year} {2014})}\BibitemShut {NoStop}%
\bibitem [{\citenamefont {Barato}\ \emph {et~al.}(2014)\citenamefont {Barato},
  \citenamefont {Hartich},\ and\ \citenamefont
  {Seifert}}]{barato2014efficiency}%
  \BibitemOpen
  \bibfield  {author} {\bibinfo {author} {\bibfnamefont {A.~C.}\ \bibnamefont
  {Barato}}, \bibinfo {author} {\bibfnamefont {D.}~\bibnamefont {Hartich}},\
  and\ \bibinfo {author} {\bibfnamefont {U.}~\bibnamefont {Seifert}},\
  }\bibfield  {title} {\bibinfo {title} {Efficiency of cellular information
  processing},\ }\href@noop {} {\bibfield  {journal} {\bibinfo  {journal} {New
  J. Phys.}\ }\textbf {\bibinfo {volume} {16}},\ \bibinfo {pages} {103024}
  (\bibinfo {year} {2014})}\BibitemShut {NoStop}%
\bibitem [{\citenamefont {Hartich}\ \emph {et~al.}(2016)\citenamefont
  {Hartich}, \citenamefont {Barato},\ and\ \citenamefont
  {Seifert}}]{hartich2016sensory}%
  \BibitemOpen
  \bibfield  {author} {\bibinfo {author} {\bibfnamefont {D.}~\bibnamefont
  {Hartich}}, \bibinfo {author} {\bibfnamefont {A.~C.}\ \bibnamefont
  {Barato}},\ and\ \bibinfo {author} {\bibfnamefont {U.}~\bibnamefont
  {Seifert}},\ }\bibfield  {title} {\bibinfo {title} {Sensory capacity: {A}n
  information theoretical measure of the performance of a sensor},\ }\href@noop
  {} {\bibfield  {journal} {\bibinfo  {journal} {Phys. Rev. E}\ }\textbf
  {\bibinfo {volume} {93}},\ \bibinfo {pages} {022116} (\bibinfo {year}
  {2016})}\BibitemShut {NoStop}%
\bibitem [{\citenamefont {Matsumoto}\ and\ \citenamefont
  {Sagawa}(2018)}]{matsumoto2018role}%
  \BibitemOpen
  \bibfield  {author} {\bibinfo {author} {\bibfnamefont {T.}~\bibnamefont
  {Matsumoto}}\ and\ \bibinfo {author} {\bibfnamefont {T.}~\bibnamefont
  {Sagawa}},\ }\bibfield  {title} {\bibinfo {title} {Role of sufficient
  statistics in stochastic thermodynamics and its implication to sensory
  adaptation},\ }\href@noop {} {\bibfield  {journal} {\bibinfo  {journal}
  {Phys. Rev. E}\ }\textbf {\bibinfo {volume} {97}},\ \bibinfo {pages} {042103}
  (\bibinfo {year} {2018})}\BibitemShut {NoStop}%
\bibitem [{\citenamefont {Schreiber}(2000)}]{schreiber2000measuring}%
  \BibitemOpen
  \bibfield  {author} {\bibinfo {author} {\bibfnamefont {T.}~\bibnamefont
  {Schreiber}},\ }\bibfield  {title} {\bibinfo {title} {Measuring information
  transfer},\ }\href@noop {} {\bibfield  {journal} {\bibinfo  {journal} {Phys.
  Rev. Lett.}\ }\textbf {\bibinfo {volume} {85}},\ \bibinfo {pages} {461}
  (\bibinfo {year} {2000})}\BibitemShut {NoStop}%
\bibitem [{\citenamefont {Biferale}\ \emph {et~al.}(2017)\citenamefont
  {Biferale}, \citenamefont {Mailybaev},\ and\ \citenamefont
  {Parisi}}]{biferale2017optimal}%
  \BibitemOpen
  \bibfield  {author} {\bibinfo {author} {\bibfnamefont {L.}~\bibnamefont
  {Biferale}}, \bibinfo {author} {\bibfnamefont {A.~A.}\ \bibnamefont
  {Mailybaev}},\ and\ \bibinfo {author} {\bibfnamefont {G.}~\bibnamefont
  {Parisi}},\ }\bibfield  {title} {\bibinfo {title} {Optimal subgrid scheme for
  shell models of turbulence},\ }\href@noop {} {\bibfield  {journal} {\bibinfo
  {journal} {Phy. Rev. E}\ }\textbf {\bibinfo {volume} {95}},\ \bibinfo {pages}
  {043108} (\bibinfo {year} {2017})}\BibitemShut {NoStop}%
\bibitem [{\citenamefont {Kolmogorov}(1962)}]{kolmogorov1962refinement}%
  \BibitemOpen
  \bibfield  {author} {\bibinfo {author} {\bibfnamefont {A.~N.}\ \bibnamefont
  {Kolmogorov}},\ }\bibfield  {title} {\bibinfo {title} {A refinement of
  previous hypotheses concerning the local structure of turbulence in a viscous
  incompressible fluid at high {R}eynolds number},\ }\href@noop {} {\bibfield
  {journal} {\bibinfo  {journal} {J. Fluid Mech.}\ }\textbf {\bibinfo {volume}
  {13}},\ \bibinfo {pages} {82} (\bibinfo {year} {1962})}\BibitemShut {NoStop}%
\bibitem [{\citenamefont {Benzi}\ \emph {et~al.}(1993)\citenamefont {Benzi},
  \citenamefont {Biferale},\ and\ \citenamefont
  {Parisi}}]{benzi1993intermittency}%
  \BibitemOpen
  \bibfield  {author} {\bibinfo {author} {\bibfnamefont {R.}~\bibnamefont
  {Benzi}}, \bibinfo {author} {\bibfnamefont {L.}~\bibnamefont {Biferale}},\
  and\ \bibinfo {author} {\bibfnamefont {G.}~\bibnamefont {Parisi}},\
  }\bibfield  {title} {\bibinfo {title} {On intermittency in a cascade model
  for turbulence},\ }\href@noop {} {\bibfield  {journal} {\bibinfo  {journal}
  {Physica D}\ }\textbf {\bibinfo {volume} {65}},\ \bibinfo {pages} {163}
  (\bibinfo {year} {1993})}\BibitemShut {NoStop}%
\bibitem [{\citenamefont {Eyink}\ \emph {et~al.}(2003)\citenamefont {Eyink},
  \citenamefont {Chen},\ and\ \citenamefont {Chen}}]{eyink2003gibbsian}%
  \BibitemOpen
  \bibfield  {author} {\bibinfo {author} {\bibfnamefont {G.~L.}\ \bibnamefont
  {Eyink}}, \bibinfo {author} {\bibfnamefont {S.}~\bibnamefont {Chen}},\ and\
  \bibinfo {author} {\bibfnamefont {Q.}~\bibnamefont {Chen}},\ }\bibfield
  {title} {\bibinfo {title} {Gibbsian hypothesis in turbulence},\ }\href@noop
  {} {\bibfield  {journal} {\bibinfo  {journal} {J. Stat. Phys.}\ }\textbf
  {\bibinfo {volume} {113}},\ \bibinfo {pages} {719} (\bibinfo {year}
  {2003})}\BibitemShut {NoStop}%
\bibitem [{\citenamefont {Chen}\ \emph {et~al.}(2003)\citenamefont {Chen},
  \citenamefont {Chen}, \citenamefont {Eyink},\ and\ \citenamefont
  {Sreenivasan}}]{chen2003kolmogorov}%
  \BibitemOpen
  \bibfield  {author} {\bibinfo {author} {\bibfnamefont {Q.}~\bibnamefont
  {Chen}}, \bibinfo {author} {\bibfnamefont {S.}~\bibnamefont {Chen}}, \bibinfo
  {author} {\bibfnamefont {G.~L.}\ \bibnamefont {Eyink}},\ and\ \bibinfo
  {author} {\bibfnamefont {K.~R.}\ \bibnamefont {Sreenivasan}},\ }\bibfield
  {title} {\bibinfo {title} {Kolmogorov’s third hypothesis and turbulent sign
  statistics},\ }\href@noop {} {\bibfield  {journal} {\bibinfo  {journal}
  {Phys. Rev. Lett.}\ }\textbf {\bibinfo {volume} {90}},\ \bibinfo {pages}
  {254501} (\bibinfo {year} {2003})}\BibitemShut {NoStop}%
\bibitem [{\citenamefont {Benzi}\ \emph {et~al.}(1998)\citenamefont {Benzi},
  \citenamefont {Biferale},\ and\ \citenamefont
  {Toschi}}]{benzi1998multiscale}%
  \BibitemOpen
  \bibfield  {author} {\bibinfo {author} {\bibfnamefont {R.}~\bibnamefont
  {Benzi}}, \bibinfo {author} {\bibfnamefont {L.}~\bibnamefont {Biferale}},\
  and\ \bibinfo {author} {\bibfnamefont {F.}~\bibnamefont {Toschi}},\
  }\bibfield  {title} {\bibinfo {title} {Multiscale velocity correlations in
  turbulence},\ }\href@noop {} {\bibfield  {journal} {\bibinfo  {journal}
  {Phys. Rev. Lett.}\ }\textbf {\bibinfo {volume} {80}},\ \bibinfo {pages}
  {3244} (\bibinfo {year} {1998})}\BibitemShut {NoStop}%
\bibitem [{\citenamefont {Pedrizzetti}\ \emph {et~al.}(1996)\citenamefont
  {Pedrizzetti}, \citenamefont {Novikov},\ and\ \citenamefont
  {Praskovsky}}]{pedrizzetti1996self}%
  \BibitemOpen
  \bibfield  {author} {\bibinfo {author} {\bibfnamefont {G.}~\bibnamefont
  {Pedrizzetti}}, \bibinfo {author} {\bibfnamefont {E.~A.}\ \bibnamefont
  {Novikov}},\ and\ \bibinfo {author} {\bibfnamefont {A.~A.}\ \bibnamefont
  {Praskovsky}},\ }\bibfield  {title} {\bibinfo {title} {Self-similarity and
  probability distributions of turbulent intermittency},\ }\href@noop {}
  {\bibfield  {journal} {\bibinfo  {journal} {Phys. Rev. E}\ }\textbf {\bibinfo
  {volume} {53}},\ \bibinfo {pages} {475} (\bibinfo {year} {1996})}\BibitemShut
  {NoStop}%
\bibitem [{\citenamefont {Chhabra}\ and\ \citenamefont
  {Sreenivasan}(1992)}]{chhabra1992scale}%
  \BibitemOpen
  \bibfield  {author} {\bibinfo {author} {\bibfnamefont {A.~B.}\ \bibnamefont
  {Chhabra}}\ and\ \bibinfo {author} {\bibfnamefont {K.}~\bibnamefont
  {Sreenivasan}},\ }\bibfield  {title} {\bibinfo {title} {Scale-invariant
  multiplier distributions in turbulence},\ }\href@noop {} {\bibfield
  {journal} {\bibinfo  {journal} {Phys. Rev. Lett.}\ }\textbf {\bibinfo
  {volume} {68}},\ \bibinfo {pages} {2762} (\bibinfo {year}
  {1992})}\BibitemShut {NoStop}%
\bibitem [{Note1()}]{Note1}%
  \BibitemOpen
  \bibinfo {note} {See Supplemental Material at [URL will be inserted by
  publisher] for details on the derivation and numerical
  simulation.}\BibitemShut {Stop}%
\bibitem [{\citenamefont {Ruelle}(1978)}]{ruelle1978inequality}%
  \BibitemOpen
  \bibfield  {author} {\bibinfo {author} {\bibfnamefont {D.}~\bibnamefont
  {Ruelle}},\ }\bibfield  {title} {\bibinfo {title} {An inequality for the
  entropy of differentiable maps},\ }\href@noop {} {\bibfield  {journal}
  {\bibinfo  {journal} {Bol. Soc. Bras. de Mat.}\ }\textbf {\bibinfo {volume}
  {9}},\ \bibinfo {pages} {83} (\bibinfo {year} {1978})}\BibitemShut {NoStop}%
\bibitem [{\citenamefont {Ito}\ and\ \citenamefont
  {Sagawa}(2015)}]{ito2015maxwell}%
  \BibitemOpen
  \bibfield  {author} {\bibinfo {author} {\bibfnamefont {S.}~\bibnamefont
  {Ito}}\ and\ \bibinfo {author} {\bibfnamefont {T.}~\bibnamefont {Sagawa}},\
  }\bibfield  {title} {\bibinfo {title} {Maxwell's demon in biochemical signal
  transduction with feedback loop},\ }\href@noop {} {\bibfield  {journal}
  {\bibinfo  {journal} {Nat. Commun.}\ }\textbf {\bibinfo {volume} {6}},\
  \bibinfo {pages} {1} (\bibinfo {year} {2015})}\BibitemShut {NoStop}%
\bibitem [{\citenamefont {Kraskov}\ \emph {et~al.}(2004)\citenamefont
  {Kraskov}, \citenamefont {St{\"o}gbauer},\ and\ \citenamefont
  {Grassberger}}]{kraskov2004estimating}%
  \BibitemOpen
  \bibfield  {author} {\bibinfo {author} {\bibfnamefont {A.}~\bibnamefont
  {Kraskov}}, \bibinfo {author} {\bibfnamefont {H.}~\bibnamefont
  {St{\"o}gbauer}},\ and\ \bibinfo {author} {\bibfnamefont {P.}~\bibnamefont
  {Grassberger}},\ }\bibfield  {title} {\bibinfo {title} {Estimating mutual
  information},\ }\href@noop {} {\bibfield  {journal} {\bibinfo  {journal}
  {Phys. Rev. E}\ }\textbf {\bibinfo {volume} {69}},\ \bibinfo {pages} {066138}
  (\bibinfo {year} {2004})}\BibitemShut {NoStop}%
\bibitem [{\citenamefont {Khan}\ \emph {et~al.}(2007)\citenamefont {Khan},
  \citenamefont {Bandyopadhyay}, \citenamefont {Ganguly}, \citenamefont
  {Saigal}, \citenamefont {Erickson~III}, \citenamefont {Protopopescu},\ and\
  \citenamefont {Ostrouchov}}]{khan2007relative}%
  \BibitemOpen
  \bibfield  {author} {\bibinfo {author} {\bibfnamefont {S.}~\bibnamefont
  {Khan}}, \bibinfo {author} {\bibfnamefont {S.}~\bibnamefont {Bandyopadhyay}},
  \bibinfo {author} {\bibfnamefont {A.~R.}\ \bibnamefont {Ganguly}}, \bibinfo
  {author} {\bibfnamefont {S.}~\bibnamefont {Saigal}}, \bibinfo {author}
  {\bibfnamefont {D.~J.}\ \bibnamefont {Erickson~III}}, \bibinfo {author}
  {\bibfnamefont {V.}~\bibnamefont {Protopopescu}},\ and\ \bibinfo {author}
  {\bibfnamefont {G.}~\bibnamefont {Ostrouchov}},\ }\bibfield  {title}
  {\bibinfo {title} {Relative performance of mutual information estimation
  methods for quantifying the dependence among short and noisy data},\
  }\href@noop {} {\bibfield  {journal} {\bibinfo  {journal} {Phys. Rev. E}\
  }\textbf {\bibinfo {volume} {76}},\ \bibinfo {pages} {026209} (\bibinfo
  {year} {2007})}\BibitemShut {NoStop}%
\bibitem [{\citenamefont {Holmes}\ and\ \citenamefont
  {Nemenman}(2019)}]{holmes2019estimation}%
  \BibitemOpen
  \bibfield  {author} {\bibinfo {author} {\bibfnamefont {C.~M.}\ \bibnamefont
  {Holmes}}\ and\ \bibinfo {author} {\bibfnamefont {I.}~\bibnamefont
  {Nemenman}},\ }\bibfield  {title} {\bibinfo {title} {Estimation of mutual
  information for real-valued data with error bars and controlled bias},\
  }\href@noop {} {\bibfield  {journal} {\bibinfo  {journal} {Phys. Rev. E}\
  }\textbf {\bibinfo {volume} {100}},\ \bibinfo {pages} {022404} (\bibinfo
  {year} {2019})}\BibitemShut {NoStop}%
\bibitem [{\citenamefont {Eyink}\ and\ \citenamefont
  {Sreenivasan}(2006)}]{Eyink_Sreenivasan}%
  \BibitemOpen
  \bibfield  {author} {\bibinfo {author} {\bibfnamefont {G.~L.}\ \bibnamefont
  {Eyink}}\ and\ \bibinfo {author} {\bibfnamefont {K.~R.}\ \bibnamefont
  {Sreenivasan}},\ }\bibfield  {title} {\bibinfo {title} {Onsager and the
  theory of hydrodynamic turbulence},\ }\href@noop {} {\bibfield  {journal}
  {\bibinfo  {journal} {Rev. Mod. Phys.}\ }\textbf {\bibinfo {volume} {78}},\
  \bibinfo {pages} {87} (\bibinfo {year} {2006})}\BibitemShut {NoStop}%
\end{thebibliography}%


\begin{thebibliography}{12}%
\makeatletter
\providecommand \@ifxundefined [1]{%
 \@ifx{#1\undefined}
}%
\providecommand \@ifnum [1]{%
 \ifnum #1\expandafter \@firstoftwo
 \else \expandafter \@secondoftwo
 \fi
}%
\providecommand \@ifx [1]{%
 \ifx #1\expandafter \@firstoftwo
 \else \expandafter \@secondoftwo
 \fi
}%
\providecommand \natexlab [1]{#1}%
\providecommand \enquote  [1]{``#1''}%
\providecommand \bibnamefont  [1]{#1}%
\providecommand \bibfnamefont [1]{#1}%
\providecommand \citenamefont [1]{#1}%
\providecommand \href@noop [0]{\@secondoftwo}%
\providecommand \href [0]{\begingroup \@sanitize@url \@href}%
\providecommand \@href[1]{\@@startlink{#1}\@@href}%
\providecommand \@@href[1]{\endgroup#1\@@endlink}%
\providecommand \@sanitize@url [0]{\catcode `\\12\catcode `\$12\catcode
  `\&12\catcode `\#12\catcode `\^12\catcode `\_12\catcode `\%12\relax}%
\providecommand \@@startlink[1]{}%
\providecommand \@@endlink[0]{}%
\providecommand \url  [0]{\begingroup\@sanitize@url \@url }%
\providecommand \@url [1]{\endgroup\@href {#1}{\urlprefix }}%
\providecommand \urlprefix  [0]{URL }%
\providecommand \Eprint [0]{\href }%
\providecommand \doibase [0]{https://doi.org/}%
\providecommand \selectlanguage [0]{\@gobble}%
\providecommand \bibinfo  [0]{\@secondoftwo}%
\providecommand \bibfield  [0]{\@secondoftwo}%
\providecommand \translation [1]{[#1]}%
\providecommand \BibitemOpen [0]{}%
\providecommand \bibitemStop [0]{}%
\providecommand \bibitemNoStop [0]{.\EOS\space}%
\providecommand \EOS [0]{\spacefactor3000\relax}%
\providecommand \BibitemShut  [1]{\csname bibitem#1\endcsname}%
\let\auto@bib@innerbib\@empty
%</preamble>
\bibitem [{\citenamefont {Tanogami}\ and\ \citenamefont
  {Araki}(2024)}]{tanogami2024information}%
  \BibitemOpen
  \bibfield  {author} {\bibinfo {author} {\bibfnamefont {T.}~\bibnamefont
  {Tanogami}}\ and\ \bibinfo {author} {\bibfnamefont {R.}~\bibnamefont
  {Araki}},\ }\bibfield  {title} {\bibinfo {title} {Information-thermodynamic
  bound on information flow in turbulent cascade},\ }\href@noop {} {\bibfield
  {journal} {\bibinfo  {journal} {Phys. Rev. Research}\ }\textbf {\bibinfo
  {volume} {6}},\ \bibinfo {pages} {013090} (\bibinfo {year}
  {2024})}\BibitemShut {NoStop}%
\bibitem [{\citenamefont {Tanogami}(2024)}]{tanogami2024scale}%
  \BibitemOpen
  \bibfield  {author} {\bibinfo {author} {\bibfnamefont {T.}~\bibnamefont
  {Tanogami}},\ }\bibfield  {title} {\bibinfo {title} {Scale locality of
  information flow in turbulence},\ }\href@noop {} {\bibfield  {journal}
  {\bibinfo  {journal} {arXiv preprint arXiv:2407.20572}\ } (\bibinfo {year}
  {2024})}\BibitemShut {NoStop}%
\bibitem [{\citenamefont {L'vov}\ \emph {et~al.}(1998)\citenamefont {L'vov},
  \citenamefont {Podivilov}, \citenamefont {Pomyalov}, \citenamefont
  {Procaccia},\ and\ \citenamefont {Vandembroucq}}]{l1998improved}%
  \BibitemOpen
  \bibfield  {author} {\bibinfo {author} {\bibfnamefont {V.~S.}\ \bibnamefont
  {L'vov}}, \bibinfo {author} {\bibfnamefont {E.}~\bibnamefont {Podivilov}},
  \bibinfo {author} {\bibfnamefont {A.}~\bibnamefont {Pomyalov}}, \bibinfo
  {author} {\bibfnamefont {I.}~\bibnamefont {Procaccia}},\ and\ \bibinfo
  {author} {\bibfnamefont {D.}~\bibnamefont {Vandembroucq}},\ }\bibfield
  {title} {\bibinfo {title} {Improved shell model of turbulence},\ }\href@noop
  {} {\bibfield  {journal} {\bibinfo  {journal} {Phys. Rev. E}\ }\textbf
  {\bibinfo {volume} {58}},\ \bibinfo {pages} {1811} (\bibinfo {year}
  {1998})}\BibitemShut {NoStop}%
\bibitem [{\citenamefont {Eyink}\ \emph {et~al.}(2003)\citenamefont {Eyink},
  \citenamefont {Chen},\ and\ \citenamefont {Chen}}]{eyink2003gibbsian}%
  \BibitemOpen
  \bibfield  {author} {\bibinfo {author} {\bibfnamefont {G.~L.}\ \bibnamefont
  {Eyink}}, \bibinfo {author} {\bibfnamefont {S.}~\bibnamefont {Chen}},\ and\
  \bibinfo {author} {\bibfnamefont {Q.}~\bibnamefont {Chen}},\ }\bibfield
  {title} {\bibinfo {title} {Gibbsian hypothesis in turbulence},\ }\href@noop
  {} {\bibfield  {journal} {\bibinfo  {journal} {J. Stat. Phys.}\ }\textbf
  {\bibinfo {volume} {113}},\ \bibinfo {pages} {719} (\bibinfo {year}
  {2003})}\BibitemShut {NoStop}%
\bibitem [{\citenamefont {Bandak}\ \emph {et~al.}(2022)\citenamefont {Bandak},
  \citenamefont {Goldenfeld}, \citenamefont {Mailybaev},\ and\ \citenamefont
  {Eyink}}]{bandak2022dissipation}%
  \BibitemOpen
  \bibfield  {author} {\bibinfo {author} {\bibfnamefont {D.}~\bibnamefont
  {Bandak}}, \bibinfo {author} {\bibfnamefont {N.}~\bibnamefont {Goldenfeld}},
  \bibinfo {author} {\bibfnamefont {A.~A.}\ \bibnamefont {Mailybaev}},\ and\
  \bibinfo {author} {\bibfnamefont {G.}~\bibnamefont {Eyink}},\ }\bibfield
  {title} {\bibinfo {title} {Dissipation-range fluid turbulence and thermal
  noise},\ }\href@noop {} {\bibfield  {journal} {\bibinfo  {journal} {Phys.
  Rev. E}\ }\textbf {\bibinfo {volume} {105}},\ \bibinfo {pages} {065113}
  (\bibinfo {year} {2022})}\BibitemShut {NoStop}%
\bibitem [{\citenamefont {Lord}\ and\ \citenamefont
  {Rougemont}(2004)}]{lord2004numerical}%
  \BibitemOpen
  \bibfield  {author} {\bibinfo {author} {\bibfnamefont {G.~J.}\ \bibnamefont
  {Lord}}\ and\ \bibinfo {author} {\bibfnamefont {J.}~\bibnamefont
  {Rougemont}},\ }\bibfield  {title} {\bibinfo {title} {A numerical scheme for
  stochastic {PDE}s with {G}evrey regularity},\ }\href@noop {} {\bibfield
  {journal} {\bibinfo  {journal} {IMA J. Numer. Anal.}\ }\textbf {\bibinfo
  {volume} {24}},\ \bibinfo {pages} {587} (\bibinfo {year} {2004})}\BibitemShut
  {NoStop}%
\bibitem [{\citenamefont {Kraskov}\ \emph {et~al.}(2004)\citenamefont
  {Kraskov}, \citenamefont {St{\"o}gbauer},\ and\ \citenamefont
  {Grassberger}}]{kraskov2004estimating}%
  \BibitemOpen
  \bibfield  {author} {\bibinfo {author} {\bibfnamefont {A.}~\bibnamefont
  {Kraskov}}, \bibinfo {author} {\bibfnamefont {H.}~\bibnamefont
  {St{\"o}gbauer}},\ and\ \bibinfo {author} {\bibfnamefont {P.}~\bibnamefont
  {Grassberger}},\ }\bibfield  {title} {\bibinfo {title} {Estimating mutual
  information},\ }\href@noop {} {\bibfield  {journal} {\bibinfo  {journal}
  {Phys. Rev. E}\ }\textbf {\bibinfo {volume} {69}},\ \bibinfo {pages} {066138}
  (\bibinfo {year} {2004})}\BibitemShut {NoStop}%
\bibitem [{\citenamefont {Khan}\ \emph {et~al.}(2007)\citenamefont {Khan},
  \citenamefont {Bandyopadhyay}, \citenamefont {Ganguly}, \citenamefont
  {Saigal}, \citenamefont {Erickson~III}, \citenamefont {Protopopescu},\ and\
  \citenamefont {Ostrouchov}}]{khan2007relative}%
  \BibitemOpen
  \bibfield  {author} {\bibinfo {author} {\bibfnamefont {S.}~\bibnamefont
  {Khan}}, \bibinfo {author} {\bibfnamefont {S.}~\bibnamefont {Bandyopadhyay}},
  \bibinfo {author} {\bibfnamefont {A.~R.}\ \bibnamefont {Ganguly}}, \bibinfo
  {author} {\bibfnamefont {S.}~\bibnamefont {Saigal}}, \bibinfo {author}
  {\bibfnamefont {D.~J.}\ \bibnamefont {Erickson~III}}, \bibinfo {author}
  {\bibfnamefont {V.}~\bibnamefont {Protopopescu}},\ and\ \bibinfo {author}
  {\bibfnamefont {G.}~\bibnamefont {Ostrouchov}},\ }\bibfield  {title}
  {\bibinfo {title} {Relative performance of mutual information estimation
  methods for quantifying the dependence among short and noisy data},\
  }\href@noop {} {\bibfield  {journal} {\bibinfo  {journal} {Phys. Rev. E}\
  }\textbf {\bibinfo {volume} {76}},\ \bibinfo {pages} {026209} (\bibinfo
  {year} {2007})}\BibitemShut {NoStop}%
\bibitem [{\citenamefont {Holmes}\ and\ \citenamefont
  {Nemenman}(2019)}]{holmes2019estimation}%
  \BibitemOpen
  \bibfield  {author} {\bibinfo {author} {\bibfnamefont {C.~M.}\ \bibnamefont
  {Holmes}}\ and\ \bibinfo {author} {\bibfnamefont {I.}~\bibnamefont
  {Nemenman}},\ }\bibfield  {title} {\bibinfo {title} {Estimation of mutual
  information for real-valued data with error bars and controlled bias},\
  }\href@noop {} {\bibfield  {journal} {\bibinfo  {journal} {Phys. Rev. E}\
  }\textbf {\bibinfo {volume} {100}},\ \bibinfo {pages} {022404} (\bibinfo
  {year} {2019})}\BibitemShut {NoStop}%
\bibitem [{\citenamefont {Kolmogorov}(1962)}]{kolmogorov1962refinement}%
  \BibitemOpen
  \bibfield  {author} {\bibinfo {author} {\bibfnamefont {A.~N.}\ \bibnamefont
  {Kolmogorov}},\ }\bibfield  {title} {\bibinfo {title} {A refinement of
  previous hypotheses concerning the local structure of turbulence in a viscous
  incompressible fluid at high {R}eynolds number},\ }\href@noop {} {\bibfield
  {journal} {\bibinfo  {journal} {J. Fluid Mech.}\ }\textbf {\bibinfo {volume}
  {13}},\ \bibinfo {pages} {82} (\bibinfo {year} {1962})}\BibitemShut {NoStop}%
\bibitem [{\citenamefont {Chen}\ \emph {et~al.}(2003)\citenamefont {Chen},
  \citenamefont {Chen}, \citenamefont {Eyink},\ and\ \citenamefont
  {Sreenivasan}}]{chen2003kolmogorov}%
  \BibitemOpen
  \bibfield  {author} {\bibinfo {author} {\bibfnamefont {Q.}~\bibnamefont
  {Chen}}, \bibinfo {author} {\bibfnamefont {S.}~\bibnamefont {Chen}}, \bibinfo
  {author} {\bibfnamefont {G.~L.}\ \bibnamefont {Eyink}},\ and\ \bibinfo
  {author} {\bibfnamefont {K.~R.}\ \bibnamefont {Sreenivasan}},\ }\bibfield
  {title} {\bibinfo {title} {Kolmogorov’s third hypothesis and turbulent sign
  statistics},\ }\href@noop {} {\bibfield  {journal} {\bibinfo  {journal}
  {Phys. Rev. Lett.}\ }\textbf {\bibinfo {volume} {90}},\ \bibinfo {pages}
  {254501} (\bibinfo {year} {2003})}\BibitemShut {NoStop}%
\bibitem [{\citenamefont {Biferale}\ \emph {et~al.}(2017)\citenamefont
  {Biferale}, \citenamefont {Mailybaev},\ and\ \citenamefont
  {Parisi}}]{biferale2017optimal}%
  \BibitemOpen
  \bibfield  {author} {\bibinfo {author} {\bibfnamefont {L.}~\bibnamefont
  {Biferale}}, \bibinfo {author} {\bibfnamefont {A.~A.}\ \bibnamefont
  {Mailybaev}},\ and\ \bibinfo {author} {\bibfnamefont {G.}~\bibnamefont
  {Parisi}},\ }\bibfield  {title} {\bibinfo {title} {Optimal subgrid scheme for
  shell models of turbulence},\ }\href@noop {} {\bibfield  {journal} {\bibinfo
  {journal} {Phy. Rev. E}\ }\textbf {\bibinfo {volume} {95}},\ \bibinfo {pages}
  {043108} (\bibinfo {year} {2017})}\BibitemShut {NoStop}%
\end{thebibliography}%

\end{document}

\onecolumngrid
% \clearpage
%%%%%%%%%% Prefix a "S" to all equations, figures, tables and reset the counter %%%%%%%%%%
\setcounter{equation}{0}
\setcounter{figure}{0}
\setcounter{table}{0}
\makeatletter
\renewcommand{\thesection}{S\arabic{section}}
\renewcommand{\theequation}{S\arabic{equation}}
\renewcommand{\thefigure}{S\arabic{figure}}
% \renewcommand{\bibnumfmt}[1]{[S#1]}
% \renewcommand{\citenumfont}[1]{S#1}
%%%%%%%%%% Prefix a "S" to all equations, figures, tables and reset the counter %%%%%%%%%%
% \appendix
\title{Supplemental Material: Scale-to-Scale Information Flow Amplifies Turbulent Fluctuations}% Force line breaks with \\

\author{Tomohiro Tanogami$^1$}
\author{Ryo Araki$^{2}$}
\affiliation{$^1$Department of Earth and Space Science, Osaka University, Osaka 560-0043, Japan\\
$^2$Department of Mechanical and Aerospace Engineering, Faculty of Science
and Technology, Tokyo University of Science, Yamazaki 2641, Noda-shi
278-8510, Japan}%Lines break automatically or can be forced with \\

% \collaboration{MUSO Collaboration}%\noaffiliation

% \affiliation{
% \affiliation{
% \affiliation{%
%  Authors' institution and/or address\\
%  This line break forced with \textbackslash\textbackslash
% }%

% \collaboration{CLEO Collaboration}%\noaffiliation

% \begin{abstract}
% \begin{description}
% Secondary publications and information retrieval purposes.
% \item[PACS numbers]
% May be entered using the \verb+\pacs{#1}+ command.
% You may use the \texttt{description} environment to structure your abstract;
% use the optional argument of the \verb+\item+ command to give the category of each item.
% \end{description}
% \end{abstract}

\pacs{Valid PACS appear here}% PACS, the Physics and Astronomy

\maketitle
\onecolumngrid
In this Supplemental Material, we provide details on the derivation and numerical simulation presented in the main text.
In Secs.~\ref{Detailed derivation of the first main result} and \ref{Detailed derivation of the second main result}, we describe a detailed derivation of the first and second main results, respectively.
% In Sec.~\ref{Detailed derivation of the second main result}, we describe a detailed derivation of the second main result.
In Sec.~\ref{Approximate calculation of C_p}, we explicitly calculate the universal constant $C_p$ by imposing the additional assumption.
In Sec.~\ref{Details of the numerical simulations}, we describe the details of the numerical simulation.
We also discuss the estimation bias of the KSG estimator.

\tableofcontents

\section{Detailed derivation of the first main result\label{Detailed derivation of the first main result}}
In this section, we provide a more detailed derivation of the first main result [Eq.~(13) in the main text].
We first note that the scale-to-scale information flow can be expressed in terms of the probability current $J_n(u,u^*)$:
\begin{align}
\dot{I}^<_K=\sum^{n_K}_{n=0}\int dudu^*\left[J_n(u,u^*)\dfrac{\partial}{\partial u_n}\ln\dfrac{p_t(u,u^*)}{p_t({\bm U}^<_K)p_t({\bm U}^>_K)}+J^*_n(u,u^*)\dfrac{\partial}{\partial u^*_n}\ln\dfrac{p_t(u,u^*)}{p_t({\bm U}^<_K)p_t({\bm U}^>_K)}\right].
\label{IF_expression_1}
\end{align}
Here, $dudu^*:=\prod_nd\mathrm{Re}[u_n]d\mathrm{Im}[u_n]$, where $\mathrm{Re}[u_n]$ and $\mathrm{Im}[u_n]$ denote the real and imaginary parts of $u_n$, respectively.
The proof of Eq.~(\ref{IF_expression_1}) is given in appendices of Refs.~\cite{tanogami2024information,tanogami2024scale}.
Note that Eq.~(\ref{IF_expression_1}) is equivalent to Eq.~(16) in the main text.

Let $K$ be within the inertial range $k_f\ll K\ll k_\nu$.
In this intermediate asymptotic limit, the viscous and thermal noise terms in $J_n(u,u^*)$ go to zero for $n\le n_K$~\cite{tanogami2024scale}:
\begin{align}
\dot{I}^<_K=\sum^{n_K}_{n=0}\int dudu^*\left[\left(B_n(u,u^*)+f_n\right)p_t(u,u^*)\dfrac{\partial}{\partial u_n}\ln\dfrac{p_t(u,u^*)}{p_t({\bm U}^<_K)p_t({\bm U}^>_K)}+\mathrm{c.c.}\right],
\end{align}
where $\mathrm{c.c.}$ denotes the complex conjugate term.
Furthermore, we can show that the external force in $J_n(u,u^*)$ does not contribute to the information flow.
In fact,
\begin{align}
&\quad\sum^{n_K}_{n=0}\int dudu^*\left[f_np_t(u,u^*)\dfrac{\partial}{\partial u_n}\ln\dfrac{p_t(u,u^*)}{p_t({\bm U}^<_K)p_t({\bm U}^>_K)}+\mathrm{c.c.}\right]\notag\\
&=\sum^{n_K}_{n=0}\int dudu^*\left[f_n\left(\dfrac{\partial}{\partial u_n}p_t(u,u^*)-p_t({\bm U}^>_K|{\bm U}^<_K)\dfrac{\partial}{\partial u_n}p_t({\bm U}^<_K)\right)+\mathrm{c.c.}\right]\notag\\
&=\sum^{n_K}_{n=0}\left[f_n\left(\int du_n\dfrac{\partial}{\partial u_n}p_t(u_n)-\int du_n\dfrac{\partial}{\partial u_n}p_t(u_n)\right)+\mathrm{c.c.}\right]\notag\\
&=0,
\end{align}
where $p_t({\bm U}^>_K|{\bm U}^<_K)=p_t(u,u^*)/p_t({\bm U}^<_K)$ denotes the conditional probability density, and $p_t(u_n)$ denotes the marginal distribution of $u_n$.
Therefore, for $K$ within the inertial range, the scale-to-scale information flow $\dot{I}^<_K$ can be expressed as
\begin{align}
\dot{I}^<_K&=\sum^{n_K}_{n=0}\int dudu^*\left[B_n(u,u^*)p_t(u,u^*)\dfrac{\partial}{\partial u_n}\ln\dfrac{p_t(u,u^*)}{p_t({\bm U}^<_K)p_t({\bm U}^>_K)}+\mathrm{c.c.}\right].
\label{IF in terms of reversible current}
\end{align}
Note that this expression implies that the scale-to-scale information flow in turbulence is governed by the nonlinear interactions rather than thermal fluctuations.
From Eq.~(\ref{IF in terms of reversible current}), it follows that
\begin{align}
\dot{I}^<_K&=\sum^{n_K}_{n=0}\int dudu^*\left[B_n(u,u^*)\dfrac{\partial}{\partial u_n}p_t(u,u^*)-B_n(u,u^*)p_t({\bm U}^>_K|{\bm U}^<_K)\dfrac{\partial}{\partial u_n}p_t({\bm U}^<_K)+\mathrm{c.c.}\right]\notag\\
&=-\sum^{n_K}_{n=0}\int dudu^*\left(\dfrac{\partial}{\partial u_n}B_n(u,u^*)+\dfrac{\partial}{\partial u^*_n}B^*_n(u,u^*)\right)p_t(u,u^*)\notag\\
&\quad +\sum^{n_K}_{n=0}\int d{\bm U}^<_K\left(\dfrac{\partial}{\partial u_n}\overline{B}_n({\bm U}^<_K)+\dfrac{\partial}{\partial u^*_n}\overline{B}^*_n({\bm U}^<_K)\right)p_t({\bm U}^<_K).
\end{align}
In the second line, we used integration by parts and the definition of the effective nonlinear term (Eq.~(8) in the main text).
By noting that the divergence of $B_n(u,u^*)$ for each shell is zero (Eq.~(11) in the main text) and by using the definition of the phase-space contraction rate $\dot{\sigma}_K$ (Eq.~(12) in the main text), we obtain
\begin{align}
\dot{I}^<_K=-\dot{\sigma}_K.
\end{align}
Since $\dot{I}^<_K=-\dot{\mathcal{I}}_K$ in the steady state, this proves Eq.~(13).

\section{Detailed derivation of the second main result\label{Detailed derivation of the second main result}}
In this section, we provide a more detailed derivation of the second main result [Eq.~(14) in the main text].
Our starting point is the equality between the information flow and the average phase-space contraction rate [Eq.~(13) in the main text], which can be restated as
\begin{align}
\dot{\mathcal{I}}_K&=\dot{\sigma}_K\notag\\
&=-\left\langle\sum^{n_K}_{n=0}\left(\dfrac{\partial}{\partial u_n}\overline{B}_n({\bm U}^<_K)+\dfrac{\partial}{\partial u^*_n}\overline{B}^*_n({\bm U}^<_K)\right)\right\rangle\quad\text{for}\quad k_f\ll K\ll k_\nu.
% :=-\sum^{n_K}_{n=0}\int d{\bm U}^<_K\left(\dfrac{\partial}{\partial u_n}\overline{B}_n({\bm U}^<_K)+\dfrac{\partial}{\partial u^*_n}\overline{B}^*_n({\bm U}^<_K)\right)p_t({\bm U}^<_K).
\label{IF = sigma}
\end{align}
The second main result can be obtained by evaluating the divergence of the effective nonlinear term $\overline{B}_n({\bm U}^<_K)$ appearing in Eq.~(\ref{IF = sigma}).

First, we rewrite $\overline{B}_n({\bm U}^<_K)$ in terms of the Kolmogorov multipliers by using several properties of the original nonlinear term $B_n(u,u^*)$.
We note that $B_n(u,u^*)$ is a homogeneous function of its arguments of degree two and possesses the following symmetry under the phase transformation $u_n\mapsto\tilde{u}_n:=u_n\exp(i\phi_n)$ with $\phi_n-\phi_{n-1}-\phi_{n-2}=0$~\cite{l1998improved}:
\begin{align}
B_n(u,u^*)=e^{-i\phi_n}B_n(\tilde{u},\tilde{u}^*).
\end{align}
This phase symmetry is analogous to the translation invariance of the Navier--Stokes equation.
Using these properties, we can rewrite $B_n(u,u^*)$ in terms of the Kolmogorov multipliers as
\begin{align}
B_n(u,u^*)=k_n|u_n|^2e^{i\theta_n}\mathcal{B}_n(z,z^*),
% \overline{B}_n({\bm U}^<_K)&=k_n|u_n|^2e^{i\theta_n}\overline{\mathcal{B}}_n({\bm Z}^{<,\mathrm{local}}_K),
% \label{effective_B_nK}
\end{align}
where $\mathcal{B}_n(z,z^*)$ is a function of the Kolmogorov multipliers $\{z,z^*\}$ defined by
\begin{align}
\mathcal{B}_n(z,z^*):=i\left(2z_{n+2}|z_{n+1}|^2-\dfrac{1}{2}\dfrac{z_{n+1}}{|z_n|}+\dfrac{1}{4}\dfrac{1}{z_n|z_n||z_{n-1}|}\right).
\end{align}
From this expression, the effective nonlinear term $\overline{B}_n({\bm U}^<_K)$ can be rewritten as follows:
\begin{align}
% B_n(u,u^*)&=k_n|u_n|^2e^{i\theta_n}\mathcal{B}_n(z,z^*),\\
\overline{B}_n({\bm U}^<_K)&:=\int d{\bm U}^>_KB_n(u,u^*)p_t({\bm U}^>_K|{\bm U}^<_K)\notag\\
&=k_n|u_n|^2e^{i\theta_n}\overline{\mathcal{B}}_n({\bm Z}^{<,\mathrm{local}}_K)\quad\text{for}\quad 0\le n\le n_K,
\label{effective_B_nK}
\end{align}
where $\overline{\mathcal{B}}_n({\bm Z}^{<,\mathrm{local}}_K)$ denotes its conditional average with respect to $p_{\mathrm{uni}}({\bm Z}^>_K|{\bm Z}^{<,\mathrm{local}}_K)$:
\begin{align}
\overline{\mathcal{B}}_n({\bm Z}^{<,\mathrm{local}}_K):=\int d{\bm Z}^>_K\mathcal{B}_n(z,z^*)p_{\mathrm{uni}}({\bm Z}^>_K|{\bm Z}^{<,\mathrm{local}}_K).
\end{align} 
Here, we used $p_t({\bm U}^>_K|{\bm U}^<_K)d{\bm U}^>_K=p_{\mathrm{uni}}({\bm Z}^>_K|{\bm Z}^{<,\mathrm{local}}_K)d{\bm Z}^>_K$ by assuming the Kolmogorov hypothesis.

We now calculate the divergence of $\overline{B}_n({\bm U}^<_K)$ appearing in Eq.~(\ref{IF = sigma}). 
For $0\le n\le n_K-2$, note that $\overline{B}_n({\bm U}^<_K)=B_n(u,u^*)$ because the direct nonlinear interaction is limited to the nearest- and next-nearest-neighbor shells, i.e., $B_n(u,u^*)$ does not depend on ${\bm U}^>_K$.
From this property, it immediately follows that
\begin{align}
\dfrac{\partial}{\partial u_{n}}\overline{B}_{n}({\bm U}^<_K)+\dfrac{\partial}{\partial u^*_{n}}\overline{B}^*_{n}({\bm U}^<_K)&=\dfrac{\partial}{\partial u_n}B_n(u,u^*)+\dfrac{\partial}{\partial u^*_n}B^*_n(u,u^*)\notag\\
&=0\quad\text{for}\quad 0\le n\le n_K-2,
\end{align}
where we used the fact that the divergence of $B_n(u,u^*)$ for each shell is zero [Eq.~(11) in the main text].
For $n=n_K-1,n_K$, from Eq.~(\ref{effective_B_nK}), the derivative of $\overline{B}_n({\bm U}^<_K)$ reads
\begin{align}
\dfrac{\partial}{\partial u_{n_K-1}}\overline{B}_{n_K-1}({\bm U}^<_K)&=\dfrac{\partial}{\partial u_{n_K-1}}\left(\dfrac{K}{2}|u_{n_K-1}|^2e^{i\theta_{n_K-1}}\overline{\mathcal{B}}_{n_K-1}({\bm Z}^{<,\mathrm{local}}_K)\right),\\
\dfrac{\partial}{\partial u_{n_K}}\overline{B}_{n_K}({\bm U}^<_K)&=\dfrac{\partial}{\partial u_{n_K}}\left(K|u_{n_K}|^2e^{i\theta_{n_K}}\overline{\mathcal{B}}_{n_K}({\bm Z}^{<,\mathrm{local}}_K)\right).
\end{align}
By noting that because $|u_n|=\sqrt{u_nu^*_n}$ and $\theta_n=(2i)^{-1}\ln(u_n/u^*_n)$,
\begin{align}
\dfrac{\partial}{\partial u_n}&=\dfrac{\partial|u_n|}{\partial u_n}\dfrac{\partial}{\partial |u_n|}+\dfrac{\partial \theta_n}{\partial u_n}\dfrac{\partial}{\partial \theta_n}\notag\\
&=\dfrac{e^{-i\theta_n}}{2}\left(\dfrac{\partial}{\partial |u_n|}-\dfrac{i}{|u_n|}\dfrac{\partial}{\partial \theta_n}\right),
\end{align}
we obtain
\begin{align}
\dfrac{\partial}{\partial u_{n_K-1}}\overline{B}_{n_K-1}({\bm U}^<_K)&=\dfrac{3}{4}K|u_{n_K-1}|\overline{\mathcal{B}}_{n_K-1}({\bm Z}^{<,\mathrm{local}}_K)+\dfrac{K}{2}|u_{n_K-1}|^2e^{i\theta_{n_K-1}}\dfrac{\partial}{\partial u_{n_K-1}}\overline{\mathcal{B}}_{n_K-1}({\bm Z}^{<,\mathrm{local}}_K),\label{calc: derivative_effective_B_nK-1}\\
\dfrac{\partial}{\partial u_{n_K}}\overline{B}_{n_K}({\bm U}^<_K)&=\dfrac{3}{2}K|u_{n_K}|\overline{\mathcal{B}}_{n_K}({\bm Z}^{<,\mathrm{local}}_K)+K|u_{n_K}|^2e^{i\theta_{n_K}}\dfrac{\partial}{\partial u_{n_K}}\overline{\mathcal{B}}_{n_K}({\bm Z}^{<,\mathrm{local}}_K).\label{calc: derivative_effective_B_nK}
\end{align}
Using the fact that $\partial z^*_n/\partial u_n=\partial z^*_{n+1}/\partial u_n=0$, the derivative of $\overline{\mathcal{B}}_{n}({\bm Z}^{<,\mathrm{local}}_K)$ in the last terms of Eqs.~(\ref{calc: derivative_effective_B_nK-1}) and (\ref{calc: derivative_effective_B_nK}) can be calculated as
\begin{align}
\dfrac{\partial}{\partial u_{n_K-1}}\overline{\mathcal{B}}_{n_K-1}({\bm Z}^{<,\mathrm{local}}_K)&=\left(\dfrac{\partial z_{n_K}}{\partial u_{n_K-1}}\dfrac{\partial}{\partial z_{n_K}}+\dfrac{\partial z_{n_K-1}}{\partial u_{n_K-1}}\dfrac{\partial}{\partial z_{n_K-1}}\right)\overline{\mathcal{B}}_{n_K-1}({\bm Z}^{<,\mathrm{local}}_K)\notag\\
&=\dfrac{e^{-i\theta_{n_K-1}}}{|u_{n_K-1}|}\left(-z_{n_K}\dfrac{\partial}{\partial z_{n_K}}+z_{n_K-1}\dfrac{\partial}{\partial z_{n_K-1}}\right)\overline{\mathcal{B}}_{n_K-1}({\bm Z}^{<,\mathrm{local}}_K),\\
\dfrac{\partial}{\partial u_{n_K}}\overline{\mathcal{B}}_{n_K}({\bm Z}^{<,\mathrm{local}}_K)&=\dfrac{\partial z_{n_K}}{\partial u_{n_K}}\dfrac{\partial}{\partial z_{n_K}}\overline{\mathcal{B}}_{n_K}({\bm Z}^{<,\mathrm{local}}_K)\notag\\
&=\dfrac{e^{-i\theta_{n_K}}}{|u_{n_K}|}z_{n_K}\dfrac{\partial}{\partial z_{n_K}}\overline{\mathcal{B}}_{n_K}({\bm Z}^{<,\mathrm{local}}_K).
\end{align}
Substituting these expressions into Eqs.~(\ref{calc: derivative_effective_B_nK-1}) and (\ref{calc: derivative_effective_B_nK}), we obtain
\begin{align}
\dfrac{\partial}{\partial u_{n_K-1}}\overline{B}_{n_K-1}({\bm U}^<_K)&=\dfrac{3}{4}K|u_{n_K-1}|\overline{\mathcal{B}}_{n_K-1}({\bm Z}^{<,\mathrm{local}}_K)\notag\\
&\quad+\dfrac{K}{2}|u_{n_K-1}|\left(-z_{n_K}\dfrac{\partial}{\partial z_{n_K}}+z_{n_K-1}\dfrac{\partial}{\partial z_{n_K-1}}\right)\overline{\mathcal{B}}_{n_K-1}({\bm Z}^{<,\mathrm{local}}_K),\\
\dfrac{\partial}{\partial u_{n_K}}\overline{B}_{n_K}({\bm U}^<_K)&=\dfrac{3}{2}K|u_{n_K}|\overline{\mathcal{B}}_{n_K}({\bm Z}^{<,\mathrm{local}}_K)+K|u_{n_K}|z_{n_K}\dfrac{\partial}{\partial z_{n_K}}\overline{\mathcal{B}}_{n_K}({\bm Z}^{<,\mathrm{local}}_K).
\end{align}
Combining the complex conjugate term, the divergence of $\overline{B}_n({\bm U}^<_K)$ for $n=n_K-1,n_K$ can be expressed as
\begin{align}
\dfrac{\partial}{\partial u_{n}}\overline{B}_{n}({\bm U}^<_K)+\dfrac{\partial}{\partial u^*_{n}}\overline{B}^*_{n}({\bm U}^<_K)=K|u_{n_K}|g_n({\bm Z}^{<,\mathrm{local}}_K)\quad&\text{for}\quad n=n_K-1,n_K,
\end{align}
% \begin{align}
% \dfrac{\partial}{\partial u_{n_K-1}}\overline{B}_{n_K-1}({\bm U}^<_K)+\dfrac{\partial}{\partial u^*_{n_K-1}}\overline{B}^*_{n_K-1}({\bm U}^<_K)=K|u_{n_K}|g_{n_K-1}({\bm Z}^{<,\mathrm{local}}_K),\\
% % &=\dfrac{3}{2}K|u_{n_K-1}|\mathrm{Re}[\overline{\mathcal{B}}_{n_K-1}({\bm Z}^{<,\mathrm{local}}_K)]\notag\\
% % &\qquad+K|u_{n_K-1}|\mathrm{Re}\left[\left(-z_{n_K}\dfrac{\partial}{\partial z_{n_K}}+z_{n_K-1}\dfrac{\partial}{\partial z_{n_K-1}}\right)\overline{\mathcal{B}}_{n_K-1}({\bm Z}^{<,\mathrm{local}}_K)\right]\notag\\
% \dfrac{\partial}{\partial u_{n_K}}\overline{B}_{n_K}({\bm U}^<_K)+\dfrac{\partial}{\partial u^*_{n_K}}\overline{B}^*_{n_K}({\bm U}^<_K)=K|u_{n_K}|g_{n_K}({\bm Z}^{<,\mathrm{local}}_K),
% % &=3K|u_{n_K}|\mathrm{Re}[\overline{\mathcal{B}}_{n_K}({\bm Z}^{<,\mathrm{local}}_K)]+2K|u_{n_K}|\mathrm{Re}\left[z_{n_K}\dfrac{\partial}{\partial z_{n_K}}\overline{\mathcal{B}}_{n_K}({\bm Z}^{<,\mathrm{local}}_K)\right]\notag\\
% \end{align}
where $g_{n_K-1}({\bm Z}^{<,\mathrm{local}}_K)$ and $g_{n_K}({\bm Z}^{<,\mathrm{local}}_K)$ are universal real-valued functions of ${\bm Z}^{<,\mathrm{local}}_K$ defined by
\begin{align}
g_{n_K-1}({\bm Z}^{<,\mathrm{local}}_K)&:=\dfrac{1}{|z_{n_K}|}\left\{\dfrac{3}{2}\mathrm{Re}[\overline{\mathcal{B}}_{n_K-1}({\bm Z}^{<,\mathrm{local}}_K)]+\mathrm{Re}\left[\left(-z_{n_K}\dfrac{\partial}{\partial z_{n_K}}+z_{n_K-1}\dfrac{\partial}{\partial z_{n_K-1}}\right)\overline{\mathcal{B}}_{n_K-1}({\bm Z}^{<,\mathrm{local}}_K)\right]\right\},\label{g_nK-1}\\
g_{n_K}({\bm Z}^{<,\mathrm{local}}_K)&:=3\mathrm{Re}[\overline{\mathcal{B}}_{n_K}({\bm Z}^{<,\mathrm{local}}_K)]+2\mathrm{Re}\left[z_{n_K}\dfrac{\partial}{\partial z_{n_K}}\overline{\mathcal{B}}_{n_K}({\bm Z}^{<,\mathrm{local}}_K)\right].\label{g_nK}
\end{align}
To summarize, the divergence of the effective nonlinear term reads
\begin{align}
\dfrac{\partial}{\partial u_{n}}\overline{B}_{n}({\bm U}^<_K)+\dfrac{\partial}{\partial u^*_{n}}\overline{B}^*_{n}({\bm U}^<_K)=
\begin{cases}
0\quad&\text{for}\quad n=0,1,\ldots,n_K-2,\\
K|u_{n_K}|g_n({\bm Z}^{<,\mathrm{local}}_K)\quad&\text{for}\quad n=n_K-1,n_K.
\end{cases}
\label{derivative of the effective nonlinear term}
\end{align}
% From Eqs.~(\ref{Liouville theorem}) and (\ref{effective_B_nK}), we can calculate the derivative of the $\overline{B}_n({\bm U}^<_K)$ as
% \begin{align}
% \dfrac{\partial}{\partial u_{n}}\overline{B}_{n}({\bm U}^<_K)+\dfrac{\partial}{\partial u^*_{n}}\overline{B}^*_{n}({\bm U}^<_K)=
% \begin{cases}
% 0\quad&\text{for}\quad n=0,1,\ldots,n_K-2,\\
% K|u_{n_K}|g_n({\bm Z}^{<,\mathrm{local}}_K)\quad&\text{for}\quad n=n_K-1,n_K,
% \end{cases}
% \label{derivative of the effective nonlinear term}
% \end{align}
% where $g_n({\bm Z}^{<,\mathrm{local}}_K)$ is a universal real-valued function of ${\bm Z}^{<,\mathrm{local}}_K$ defined by
% \begin{align}
% g_{n_K-1}({\bm Z}^{<,\mathrm{local}}_K)&:=\dfrac{1}{|z_{n_K}|}\left\{\dfrac{3}{2}\mathrm{Re}[\overline{\mathcal{B}}_{n_K-1}({\bm Z}^{<,\mathrm{local}}_K)]+\mathrm{Re}\left[\left(-z_{n_K}\dfrac{\partial}{\partial z_{n_K}}+z_{n_K-1}\dfrac{\partial}{\partial z_{n_K-1}}\right)\overline{\mathcal{B}}_{n_K-1}({\bm Z}^{<,\mathrm{local}}_K)\right]\right\},\label{g_nK-1}\\
% g_{n_K}({\bm Z}^{<,\mathrm{local}}_K)&:=3\mathrm{Re}[\overline{\mathcal{B}}_{n_K}({\bm Z}^{<,\mathrm{local}}_K)]+2\mathrm{Re}\left[z_{n_K}\dfrac{\partial}{\partial z_{n_K}}\overline{\mathcal{B}}_{n_K}({\bm Z}^{<,\mathrm{local}}_K)\right].\label{g_nK}
% \end{align}

By substituting Eq.~(\ref{derivative of the effective nonlinear term}) into Eq.~(\ref{IF = sigma}), we obtain
\begin{align}
\dot{\mathcal{I}}_K=K\biggl\langle|u_{n_K}|G({\bm Z}^{<,\mathrm{local}}_K)\biggr\rangle\qquad \text{for}\quad k_f\ll K\ll k_\nu,
\label{IF equality}
\end{align}
where $G({\bm Z}^{<,\mathrm{local}}_K)$ is a universal real-valued function of ${\bm Z}^{<,\mathrm{local}}_K$ defined by
\begin{align}
G({\bm Z}^{<,\mathrm{local}}_K):=-g_{n_K-1}({\bm Z}^{<,\mathrm{local}}_K)-g_{n_K}({\bm Z}^{<,\mathrm{local}}_K).
\label{def: G}
\end{align}
By using the H\"older inequality, we finally obtain
\begin{align}
\dot{\mathcal{I}}_K\le C_pK\left\langle|u_{n_K}|^p\right\rangle^{1/p}\qquad \text{for}\quad p\in[1,\infty],
% \label{IF upper bound}
\end{align}
in the inertial range.
Here, $C_p$ is defined by
\begin{align}
C_p:=\left\langle|G({\bm Z}^{<,\mathrm{local}}_K)|^q\right\rangle^{1/q}\quad\text{with}\quad\dfrac{1}{p}+\dfrac{1}{q}=1,
\label{def C_p}
\end{align}
where $\langle\cdot\rangle$ denotes the average with respect to the probability density $p_{\mathrm{uni}}({\bm Z}^{<,\mathrm{local}}_K))$, which is also universal and time-independent.
Note that, from the Kolmogorov hypothesis, $C_p$ is a universal constant and independent of $K$.

\section{Approximate calculation of $C_p$\label{Approximate calculation of C_p}}
In this section, we explicitly calculate the universal constant $C_p$ by imposing the additional assumption of the statistical independence of the Kolmogorov multipliers.
First, we note that under this assumption, the probability distribution $p_{\mathrm{uni}}(z_n)$ for the Kolmogorov multiplier $z_n=|z_n|e^{i\Delta_n}$ is given by (for the derivation, see Appendix of~\cite{eyink2003gibbsian})
\begin{align}
p_{\mathrm{uni}}(z_n)=\delta(z_{n}-\overline{z}_n),
\label{delta function}
\end{align}
where $\overline{z}_n=|\overline{z}_n|e^{i\overline{\Delta}_n}=2^{-1/3}i$ ($|\overline{z}_n|:=2^{-1/3}$ and $\overline{\Delta}_n:=\pi/2$) corresponds to the (unstable) K41 solution of the Sabra shell model of the form $u_n=-iAk^{-1/3}_n$ with an arbitrary constant $A>0$.
% We write the corresponding multiplier as $\overline{z}_n=|\overline{z}_n|e^{i\overline{\Delta}_n}=2^{-1/3}i$, i.e., $|\overline{z}_n|:=2^{-1/3}$ and $\overline{\Delta}_n:=\pi/2$.
Because $z_n$ is assumed to be statistically independent, the probability density for the multipliers, such as $p_{\mathrm{uni}}({\bm Z}^{<,\mathrm{local}}_K))$ and $p_{\mathrm{uni}}({\bm Z}^>_K|{\bm Z}^{<,\mathrm{local}}_K)$, is a product of Eq.~(\ref{delta function}).
% the conditional probability distribution $p_{\mathrm{uni}}(z_{n_K+2},z^*_{n_K+2},z_{n_K+1},z^*_{n_K+1}|{\bm Z}^{<,\mathrm{local}}_K)$ is also given by
% \begin{align}
% p_{\mathrm{uni}}(z_{n_K+2},z^*_{n_K+2},z_{n_K+1},z^*_{n_K+1}|{\bm Z}^{<,\mathrm{local}}_K)=\delta(z_{n_K+2}-2^{-1/3}i)\delta(z^*_{n_K+2}+2^{-1/3}i)\delta(z_{n_K+1}-2^{-1/3}i)\delta(z^*_{n_K+1}+2^{-1/3}i).
% % =\delta(w_{n_K+2}-2^{-1/3})\delta\left(\Delta_{n_K+2}-\dfrac{\pi}{2}\right)\delta(w_{n_K+1}-2^{-1/3})\delta\left(\Delta_{n_K+1}-\dfrac{\pi}{2}\right).
% \label{ansatz}
% \end{align}
Using these properties, $C_p$ can be expressed as
\begin{align}
C_p&:=\left\langle|G({\bm Z}^{<,\mathrm{local}}_K)|^q\right\rangle^{1/q}\notag\\
&=\left.|G({\bm Z}^{<,\mathrm{local}}_K)|\right|_{z,z^*=\overline{z},\overline{z}^*},
% \label{def C_p}
\end{align}
where $\left.\cdot\right|_{z,z^*=\overline{z},\overline{z}^*}$ means substituting $\overline{z},\overline{z}^*$ for $z,z^*$.
Recalling the definition of $G({\bm Z}^{<,\mathrm{local}}_K)$ [Eq.~(\ref{def: G})], we now calculate the values of $g_{n_K-1}({\bm Z}^{<,\mathrm{local}}_K)$ [Eq.~(\ref{g_nK-1})] and $g_{n_K}({\bm Z}^{<,\mathrm{local}}_K)$ [Eq.~(\ref{g_nK})] when $z,z^*=\overline{z},\overline{z}^*$.
First, we note that
\begin{align}
\bigl.\mathcal{B}_n(z,z^*)\bigr|_{z,z^*=\overline{z},\overline{z}^*}&=i\left(2\overline{z}_{n+2}|\overline{z}_{n+1}|^2-\dfrac{1}{2}\dfrac{\overline{z}_{n+1}}{|\overline{z}_n|}+\dfrac{1}{4}\dfrac{1}{\overline{z}_n|\overline{z}_n||\overline{z}_{n-1}|}\right)\notag\\
&=0.
\end{align}
Hence, we can easily see that
\begin{align}
\left.\overline{\mathcal{B}}_{n_K-1}({\bm Z}^{<,\mathrm{local}}_K)\right|_{z,z^*=\overline{z},\overline{z}^*}&=0,\\
\left.\overline{\mathcal{B}}_{n_K}({\bm Z}^{<,\mathrm{local}}_K)\right|_{z,z^*=\overline{z},\overline{z}^*}&=0,
\end{align}
and therefore, the first term on the right-hand side of Eqs.~(\ref{g_nK-1}) and (\ref{g_nK}) vanishes.
For the remaining terms on the right-hand side of Eqs.~(\ref{g_nK-1}) and (\ref{g_nK}), we first note that
\begin{align}
\overline{\mathcal{B}}_{n_K-1}({\bm Z}^{<,\mathrm{local}}_K)&=i\left(2\left.\bigl\langle z_{n_K+1}\right|{\bm Z}^{<,\mathrm{local}}_K\bigr\rangle|z_{n_K}|^2-\dfrac{1}{2}\dfrac{z_{n_K}}{|z_{n_K-1}|}+\dfrac{1}{4}\dfrac{1}{z_{n_K-1}|z_{n_K-1}||z_{n_K-2}|}\right)\notag\\
&=i\left(2\overline{z}_{n_K+1}|z_{n_K}|^2-\dfrac{1}{2}\dfrac{z_{n_K}}{|z_{n_K-1}|}+\dfrac{1}{4}\dfrac{1}{z_{n_K-1}|z_{n_K-1}||z_{n_K-2}|}\right),\\
\overline{\mathcal{B}}_{n_K}({\bm Z}^{<,\mathrm{local}}_K)&=i\left(2\left.\bigl\langle z_{n_K+2}|z_{n_K+1}|^2\right|{\bm Z}^{<,\mathrm{local}}_K\bigr\rangle-\dfrac{1}{2}\dfrac{\left.\bigl\langle z_{n_K+1}\right|{\bm Z}^{<,\mathrm{local}}_K\bigr\rangle}{|z_{n_K}|}+\dfrac{1}{4}\dfrac{1}{z_{n_K}|z_{n_K}||z_{n_K-1}|}\right)\notag\\
&=i\left(2\overline{z}_{n_K+2}|\overline{z}_{n_K+1}|^2-\dfrac{1}{2}\dfrac{\overline{z}_{n_K+1}}{|z_{n_K}|}+\dfrac{1}{4}\dfrac{1}{z_{n_K}|z_{n_K}||z_{n_K-1}|}\right),
\end{align}
where $\langle\cdot|{\bm Z}^{<,\mathrm{local}}_K\rangle$ denotes the average with respect to $p_{\mathrm{uni}}({\bm Z}^>_K|{\bm Z}^{<,\mathrm{local}}_K)$, which is a product of Eq.~(\ref{delta function}) under the additional assumption.
Then, we find that
\begin{align}
&\quad\left.\left(-z_{n_K}\dfrac{\partial}{\partial z_{n_K}}+z_{n_K-1}\dfrac{\partial}{\partial z_{n_K-1}}\right)\overline{\mathcal{B}}_{n_K-1}({\bm Z}^{<,\mathrm{local}}_K)\right|_{z,z^*=\overline{z},\overline{z}^*}\notag\\
&=\left.\left[-z_{n_K}\dfrac{e^{-i\Delta_{n_K}}}{2}\left(\dfrac{\partial}{\partial |z_{n_K}|}-\dfrac{i}{|z_{n_K}|}\dfrac{\partial}{\partial \Delta_{n_K}}\right)\right.\right.\notag\\
&\qquad\left.\left.+z_{n_K-1}\dfrac{e^{-i\Delta_{n_K-1}}}{2}\left(\dfrac{\partial}{\partial |z_{n_K-1}|}-\dfrac{i}{|z_{n_K-1}|}\dfrac{\partial}{\partial \Delta_{n_K-1}}\right)\right]\overline{\mathcal{B}}_{n_K-1}({\bm Z}^{<,\mathrm{local}}_K)\right|_{z,z^*=\overline{z},\overline{z}^*}\notag\\
&=-\overline{z}_{n_K}i\left(2\overline{z}_{n_K+1}|\overline{z}_{n_K}|e^{-i\overline{\Delta}_{n_K}}-\dfrac{1}{2|\overline{z}_{n_K-1}|}\right)+\overline{z}_{n_K-1}i\left(\dfrac{e^{-i\overline{\Delta}_{n_K-1}}}{4}\dfrac{\overline{z}_{n_K}}{|\overline{z}_{n_K-1}|^2}-\dfrac{3}{4}\dfrac{e^{-i\overline{\Delta}_{n_K-1}}}{2}\dfrac{e^{-i\overline{\Delta}_{n_K-1}}}{|\overline{z}_{n_K-1}|^3|\overline{z}_{n_K-2}|}\right)\notag\\
&=-\dfrac{1}{2},
\end{align}
and
\begin{align}
\left.z_{n_K}\dfrac{\partial}{\partial z_{n_K}}\overline{\mathcal{B}}_{n_K}({\bm Z}^{<,\mathrm{local}}_K)\right|_{z,z^*=\overline{z},\overline{z}^*}&=\left.z_{n_K}\dfrac{e^{-i\Delta_{n_K}}}{2}\left(\dfrac{\partial}{\partial |z_{n_K}|}-\dfrac{i}{|z_{n_K}|}\dfrac{\partial}{\partial \Delta_{n_K}}\right)\overline{\mathcal{B}}_{n_K}({\bm Z}^{<,\mathrm{local}}_K)\right|_{z,z^*=\overline{z},\overline{z}^*}\notag\\
&=\overline{z}_{n_K}i\left(\dfrac{e^{-i\overline{\Delta}_{n_K}}}{4}\dfrac{\overline{z}_{n_K+1}}{|\overline{z}_{n_K}|^2}-\dfrac{3}{4}\dfrac{e^{-i\overline{\Delta}_{n_K}}}{2}\dfrac{e^{-i\overline{\Delta}_{n_K}}}{|\overline{z}_{n_K}|^3|\overline{z}_{n_K-1}|}\right)\notag\\
&=-1.
\end{align}
By combining these results, we obtain
\begin{align}
\left.G({\bm Z}^{<,\mathrm{local}}_K)\right|_{z,z^*=\overline{z},\overline{z}^*}&=-\left.g_{n_K}({\bm Z}^{<,\mathrm{local}}_K)\right|_{z,z^*=\overline{z},\overline{z}^*}-\left.g_{n_K-1}({\bm Z}^{<,\mathrm{local}}_K)\right|_{z,z^*=\overline{z},\overline{z}^*}\notag\\
&=2+2^{-2/3}.
\end{align}
Therefore, we conclude that $C_p$ becomes independent of $p$ under the additional assumption of the statistical independence of the Kolmogorov multipliers:
\begin{align}
C_p=2+2^{-2/3}=:C.
\end{align}
Moreover, from the relation (\ref{IF equality}), we find that 
\begin{align}
\dot{\mathcal{I}}_K=CK\langle|u_{n_K}|\rangle=\dfrac{C}{\tau_K}\qquad \text{for}\quad k_f\ll K\ll k_\nu,
\end{align}
where $\tau_K:=(K\langle|u_{n_K}|\rangle)^{-1}$ denotes the eddy turnover time at scale $K$.
Thus, under the additional assumption, the equality of Eq.~(14) in the main text is achieved for $p=1$.

\section{Details of the numerical simulations\label{Details of the numerical simulations}}
In this section, we describe the details of the numerical simulation.
The setup of the numerical simulation is only briefly described because it is the same as that in our previous paper \cite{tanogami2024information}.
Then, we explain the definition of the scale-local information flow, which is estimated in the numerical simulation.
Finally, we discuss the estimation bias of the KSG estimator.

\subsection{Setup}
We numerically solve the stochastic Sabra shell model nondimensionalized by the dissipation scale $k_\nu$ and the velocity scale $u_\nu:=(\epsilon\nu)^{1/4}$:
\begin{align}
\partial_{t}{u}_n&=B_n(u,u^*)-k^2_nu_n+(2\theta_\nu)^{1/2}k_n\xi_n+\mathcal{F}_\nu f_n.
\label{sabra shell model_dimensionless}
\end{align}
Here, $\theta_\nu:=k_{\mathrm{B}}T/\rho u^2_\nu$ denotes the dimensionless temperature, and $\mathcal{F}_\nu:=F/(k_\nu u^2_\nu)$ denotes the nondimensionalized magnitude of the force.
In this case, the unit of time is given by the characteristic time scale at the dissipation scale: $\tau_\nu:=(k_\nu u_\nu)^{-1}$.

We have performed numerical simulations for two cases.
In the first case, we choose the parameter values appropriate for the atmospheric boundary layer.
Specifically, we set $N=22$, $n_f=1$, $\theta_\nu=2.328\times10^{-8}$, and
\begin{align}
\mathcal{F}_\nu f_0&=-0.008900918232183095- 0.0305497603210104i,\\
\mathcal{F}_\nu f_1&=0.005116337459331228- 0.018175040700335127i,
\end{align}
following Refs.~\cite{bandak2022dissipation,tanogami2024information}.
In this case, the achieved Reynolds number is comparable to the typical values in the atmospheric boundary layer, i.e., $\mathrm{Re}\sim10^6$.
In the second case, we set $N=19$, $\theta_\nu=2.328\times10^{-8}$, $n_f=1$, and
\begin{align}
\mathcal{F}_\nu f_0&=-0.017415685046854878- 0.05977417049893835i,\\
\mathcal{F}_\nu f_1&=0.010010711194151034- 0.03556158772544649i.
\end{align}
In this case, the achieved Reynolds number is lowered to $10^5$.
In both cases, we use a slaved $3/2$-strong-order Ito-Taylor scheme~\cite{lord2004numerical} with time-step $\delta t:=10^{-5}$, which is smaller than the viscous time scale at the highest wave number.

We estimate the mutual information by using the KSG estimator~\cite{kraskov2004estimating,khan2007relative,holmes2019estimation}, which has the advantage that it does not require estimation of the underlying probability density.
The KSG estimator uses the distances to the $\kappa$-th nearest neighbors of the sample points in the data to detect the structures of the underlying probability distribution, where $\kappa\in\mathbb{N}$ denotes the parameter of the KSG estimator.
Specifically, for the mutual information $I[X:Y]$ (either or both of the random variables $X$ and $Y$ can be multidimensional) is defined as follows~\cite{kraskov2004estimating}:
\begin{align}
\hat{I}^{(\kappa)}_{\mathrm{KSG}}[X\colon\!Y]&:=\psi(\kappa)-\dfrac{1}{\kappa}+\psi(N_{\mathrm{samp}})-\dfrac{1}{N_{\mathrm{samp}}}\sum^{N_{\mathrm{samp}}}_{i=1}\left[\psi(n_x(i))+\psi(n_y(i))\right],
\label{KSG estimator}
\end{align}
where $\psi$ is the digamma function, $N_{\mathrm{samp}}$ denotes the total number of samples, and $n^{(\kappa)}_\alpha(i)$ ($\alpha=x,y$) is the number of samples such that $\|\alpha_j-\alpha_i\|\le\epsilon^{(\kappa)}_\alpha(i)/2$.
Here, $\epsilon^{(\kappa)}_\alpha(i)$ denotes the $\alpha$ extent of the smallest hyper-rectangle in the $(x,y)$ space centered at the $i$-th sample $(x_i,y_i)$ that contains $\kappa$ of its neighboring samples.
While any norm can be used for $\|\alpha_j-\alpha_i\|$, we use the standard Euclidean norm here.

In the numerical simulation, we used $N_{\mathrm{samp}}=3\times10^5$ samples.
In the first case, these samples were obtained by sampling $100$ snapshots at time $t=1000i$ ($i=1,2,\cdots,100$) for each of the $3000$ noise realizations.
That is, for each of the $3000$ independent runs, we sampled $100$ snapshots.
Here, the time interval of the sampling, $1000$, was chosen to be larger than one large-eddy turnover time $\tau_L/\tau_\nu\simeq734<1000$.
Similarly, for the second case, $N_{\mathrm{samp}}=3\times10^5$ samples are obtained by sampling $100$ snapshots at time $t=500i$ ($i=1,2,\cdots,100$) for each of the $3000$ noise realizations, where the time interval, $500$, is chosen to be larger than one large-eddy turnover time $\tau_L/\tau_\nu\simeq181<500$.

\subsection{Definition of scale-local information flow}
\begin{figure}[b]
\center
\includegraphics[width=8.6cm]{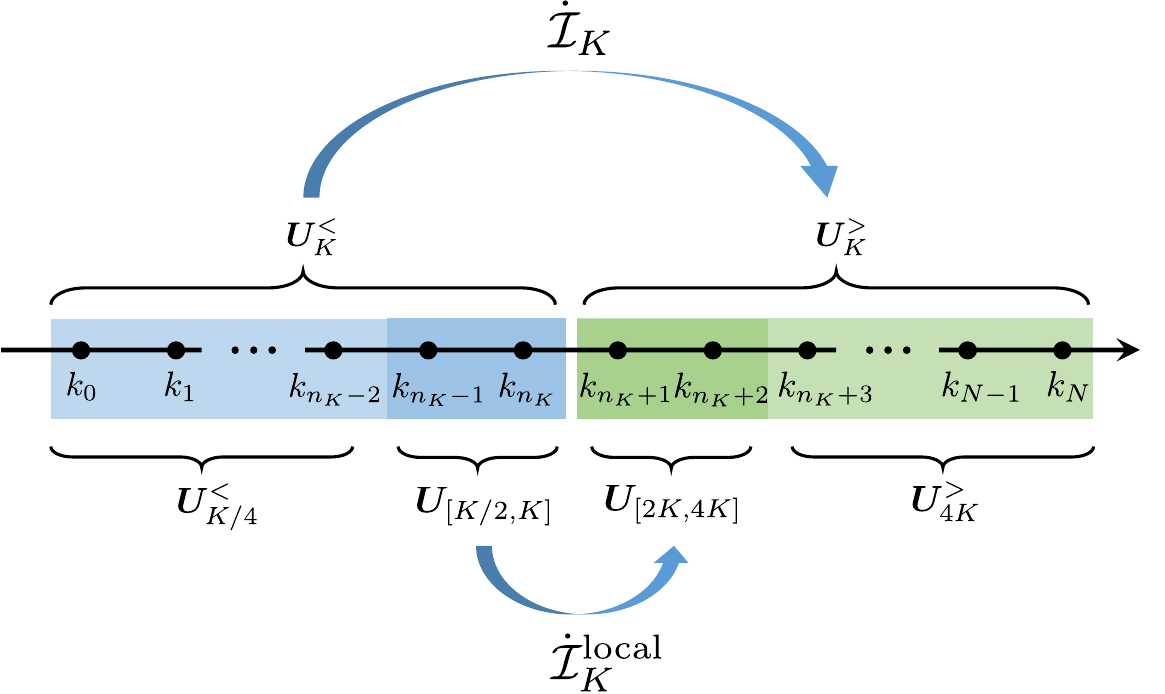}
\caption{Schematic of band-pass filtered shell variables and the scale-local information flow $\dot{\mathcal{I}}^{\mathrm{local}}_K$.
${\bm U}^<_{K/4}:=\{u_n,u^*_n\mid0\le n\le n_K-2\}$ and ${\bm U}^>_{4K}:=\{u_n,u^*_n\mid n_K+2<n\le N\}$ denote the low-pass and high-pass filtered shell variables.
}
\label{fig:Fourier_modes_division_suppl}
\end{figure}

Because it is numerically demanding to estimate $\dot{\mathcal{I}}_K$ with high precision, we instead estimate the scale-local information flow $\dot{\mathcal{I}}^{\mathrm{local}}_K$ introduced in Ref.~\cite{tanogami2024scale}.
Although the nature of $\dot{\mathcal{I}}^{\mathrm{local}}_K$ is thoroughly investigated in Ref.~\cite{tanogami2024scale}, here we briefly review its definition and relevant properties.
First, we define the scale-local modes ${\bm U}_{[K/2,K]}$ and ${\bm U}_{[2K,4K]}$ as band-pass filtered shell variables as follows (see Fig.~\ref{fig:Fourier_modes_division_suppl}):
\begin{align}
{\bm U}_{[K/2,K]}&:=\{u_{n_K-1},u^*_{n_K-1},u_{n_K},u^*_{n_K}\},\\
{\bm U}_{[2K,4K]}&:=\{u_{n_K+1},u^*_{n_K+1},u_{n_K+2},u^*_{n_K+2}\}.
\end{align}
Here, note that $K/2=k_{n_K-1}$, $K=k_{n_K}$, $2K=k_{n_K+1}$, and $4K=k_{n_K+2}$.
Then, $\dot{\mathcal{I}}^{\mathrm{local}}_K$ is defined as the steady-state value of the scale-local information flow $\dot{I}_{[2K,4K]}[{\bm U}_{[K/2,K]}\colon\!{\bm U}_{[2K,4K]}]$ defined by
\begin{align}
% \dot{\mathcal{I}}^{\mathrm{local}}_K&\equiv
\dot{I}_{[2K,4K]}[{\bm U}_{[K/2,K]}\colon\!{\bm U}_{[2K,4K]}]&:=\lim_{dt\rightarrow0^+}\dfrac{1}{dt}\biggl(I[{\bm U}_{[K/2,K]}(t)\colon\!{\bm U}_{[2K,4K]}(t+dt)]-I[{\bm U}_{[K/2,K]}(t)\colon\!{\bm U}_{[2K,4K]}(t)]\biggr),
% &=\sum^{n_K+2}_{n=n_K+1}\int dudu^*\left[J_n(u,u^*)\dfrac{\partial}{\partial u_n}\ln\dfrac{p_t({\bm U}_{[K/2,K]},{\bm U}_{[2K,4K]})}{p_t({\bm U}_{[K/2,K]})p_t({\bm U}_{[2K,4K]})}+\mathrm{c.c.}\right],
% \label{scale-local information flow}
\end{align}
which quantifies the information flow from ${\bm U}_{[K/2,K]}$ to ${\bm U}_{[2K,4K]}$ (see Fig.~\ref{fig:Fourier_modes_division_suppl}).
In Ref.~\cite{tanogami2024scale}, we proved that $\dot{\mathcal{I}}_K$ is approximately equal to $\dot{\mathcal{I}}^{\mathrm{local}}_K$ for $K$ within the inertial range:
\begin{align}
\dot{\mathcal{I}}_K\simeq\dot{\mathcal{I}}^{\mathrm{local}}_K\qquad \text{for}\quad k_f\ll K\ll k_\nu.
\label{scale locality}
\end{align}
Here, we used the approximation that the conditional probability density $p_t({\bm U}^>_{4K}|{\bm U}^<_{4K})$ has a short-range dependence on the large-scale modes ${\bm U}^<_{4K}$, i.e., $p_t({\bm U}^>_{4K}|{\bm U}^<_{4K})\simeq p_t({\bm U}^>_{4K}|{\bm U}_{[K/2,4K]})$, where ${\bm U}_{[K/2,4K]}={\bm U}_{[K/2,K]}\cup{\bm U}_{[2K,4K]}$.
This approximation is based on Kolmogorov's hypothesis~\cite{kolmogorov1962refinement,chen2003kolmogorov,eyink2003gibbsian}, and its validity is verified numerically in Ref.~\cite{biferale2017optimal}.

In estimating $\dot{\mathcal{I}}^{\mathrm{local}}_K$, we first estimate the mutual informations $I[{\bm U}_{[K/2,K]}(t+\Delta t):{\bm U}_{[2K,4K]}(t)]$ and $I[{\bm U}_{[K/2,K]}(t):{\bm U}_{[2K,4K]}(t)]$ for some suitable time increment $\Delta t$ in the steady state by using the KSG estimator.
Then, we estimate $\dot{\mathcal{I}}^{\mathrm{local}}_K$ by using a finite difference approximation:
\begin{align}
-\dfrac{\Delta I}{\Delta t}:=-\dfrac{1}{\Delta t}\biggl(&I[{\bm U}_{[K/2,K]}(t+\Delta t):{\bm U}_{[2K,4K]}(t)]-I[{\bm U}_{[K/2,K]}(t):{\bm U}_{[2K,4K]}(t)]\biggr).
% -\dfrac{I[{\bm U}^<_K(t+\Delta t):{\bm U}^>_K(t)]-I[{\bm U}^<_K(t):{\bm U}^>_K(t)]}{\Delta t}.
\label{estimated IF}
\end{align}
Note that the estimation error is amplified for a smaller $\Delta t$.
Since we are interested in the behavior of the information flow in the inertial range, we choose $\Delta t=0.1\tau_\nu$.
Consequently, we cannot resolve the scale dependence of the information flow in the dissipation range $\gtrsim k_\nu$.

\begin{figure}[t]
\center
\includegraphics[width=8.6cm]{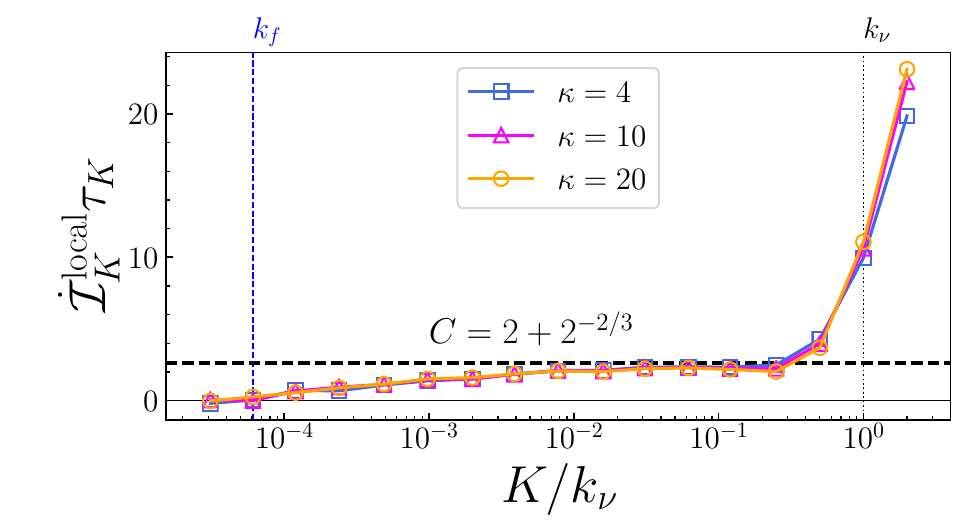}
\caption{Scale dependence of the scale-to-scale information flow $\dot{\mathcal{I}}^{\mathrm{local}}_K$ scaled by the eddy turnover time $\tau_K$ for various values of $\kappa$.
The horizontal dashed line represents $C=2+2^{-2/3}$.
The vertical dotted and dashed lines represent the dissipation scale $k_\nu$ and injection scale $k_f$, respectively.
}
\label{fig:bias}
\end{figure}

\subsection{Estimation bias of the KSG estimator}
Note that $\kappa$ is the only free parameter of the KSG estimator.
By varying $\kappa$, we can detect the structure of the underlying probability distribution in different spatial resolutions.
Although the KSG estimator is asymptotically unbiased for sufficiently regular probability distributions as $N_{\mathrm{samp}}\rightarrow\infty$, there is a $\kappa$-dependent bias for a finite $N_{\mathrm{samp}}$~\cite{holmes2019estimation}.
While $\kappa=4$ is used in Fig.~2 of the main text following Ref.~\cite{kraskov2004estimating}, here we show the estimation results for other values of $\kappa$.
Figure~\ref{fig:bias} shows the scale-to-scale information flow $\dot{\mathcal{I}}^{\mathrm{local}}_K$ scaled by the eddy turnover time $\tau_K$ for $\kappa=4, 10, 20$.
This figure suggests that the estimated information flow is independent of $\kappa$ in the inertial range. 
This result is in contrast to our previous result in Ref.~\cite{tanogami2024information}, where the estimated information flow $\dot{\mathcal{I}}_K$ (not $\dot{\mathcal{I}}^{\mathrm{local}}_K$) significantly depends on $\kappa$.

% \newpage
\bibliography{main_text}